\documentclass[11pt,a4paper]{article}
\usepackage{t1enc}
\usepackage[latin1]{inputenc}
\usepackage[english]{babel}
\normalfont
\usepackage{amsmath}
\usepackage{yfonts}
\usepackage{bbm}
\usepackage{bm}
\usepackage{wasysym}
\usepackage{amssymb}
\usepackage{mathrsfs}
\usepackage{pifont}
\usepackage{hyperref}

\newcommand{\be}[0]{\begin{equation}}
\newcommand{\ee}[0]{\end{equation}}
\renewcommand\arraystretch{1.5}

\renewcommand{\arraystretch}{.8}
\numberwithin{equation}{section}

\begin{document}

\vspace*{-1cm}
\thispagestyle{empty}
\begin{flushright}
LPTENS 11/40
\end{flushright}
\vspace*{1.5cm}

\begin{center}
{\Large 
{\bf The global gravitational anomaly of the self-dual field theory}}
\vspace{2.0cm}

{\large Samuel Monnier}
\vspace*{0.5cm}

Laboratoire de Physique Th\'eorique de l'\'Ecole 
Normale Sup\'erieure\footnote{Unit\'e mixte de recherche (UMR 8549) du CNRS  et de l'ENS, associ\'ee \`a l'Universit\'e  Pierre et Marie Curie et aux f\'ed\'erations de recherche FR684 et FR2687.} 
\\
CNRS UMR 8549 \\
24 rue Lhomond, 75231 Paris Cedex 05, France \\ \vspace{.5cm}

Laboratoire de Physique Théorique et Hautes \'Energies \\
CNRS UMR 7589 and Université Pierre et Marie Curie - Paris 6 \\
4 place Jussieu, 75252 Paris cedex 05, France\\ \vspace{.5cm}
monnier@lpt.ens.fr 
\\ 

\vspace*{1cm}

{\bf Abstract}
\end{center}

We derive a formula for the global gravitational anomaly of the self-dual field theory on an arbitrary compact oriented Riemannian manifold. Along the way, we uncover interesting links between the theory of determinant line bundles of Dirac operators, Siegel theta functions and a functor constructed by Hopkins and Singer. We apply our result to type IIB supergravity and show that in the naive approximation where the Ramond-Ramond fields are treated as differential cohomology classes, the global gravitational anomaly vanishes on all 10-dimensional spin manifolds. We sketch a few other important physical applications.

\newpage

\tableofcontents

\section{Introduction and summary}

In this paper, we continue a program started in \cite{Monnier2011} and aiming at the determination of the global gravitational anomaly of the self-dual field theory. We refer the reader to the introduction of \cite{Monnier2011} for some background on gravitational anomalies. In \cite{Monnier2011}, we managed to characterize the topological class of the anomaly bundle of the self-dual field theory, the ``topological anomaly''. In the current paper, we derive a formula for the holonomies of the natural connection living on the anomaly bundle, i.e. the global gravitational anomaly. We do not consider the problem completely solved, because the anomaly formula involves a quantity for which we do not have a general construction. In particular, a number of issues remain open for the self-dual field in dimension six.

The self-dual field theory is a central piece in the web linking string theories, M-theory and supergravities. The knowledge of its global gravitational anomaly should impact several topics. Witten's original motivation to compute it was to check anomaly cancellation in type IIB supergravity \cite{Witten:1985xe}, whose Ramond-Ramond four-form gauge field is self-dual. However, the formula he derived could only be trusted in the case when the ten dimensional spin manifold on which the theory is defined has no cohomology in degree five. In Section \ref{SecGlobAnCan}, we perform again his computation using our anomaly formula and show that the global anomaly vanishes on all 10-dimensional spin manifolds. It should be stressed that as it stands, this check is slightly naive, because it ignores the fact that Ramond-Ramond fields are differential K-theory classes \cite{Moore:1999gb, Diaconescu:2000wy, Evslin:2006cj}, which could modify the instanton sum of the self-dual field. We believe that taking the K-theoretic nature of the type IIB self-dual field into account will not radically change the anomaly formula, but we will not attempt to do so here. 

Applications to six dimensions are probably a bit more remote, as there are still some issues to understand in this case. They are nevertheless of great importance. The self-dual field appears on the M5-brane, as well as on the type IIA NS5-brane. The anomaly formula should allow to check global gravitational anomaly cancellations in M-theory \cite{Henningson:1997da, Becker:1999kh} and IIA backgrounds containing five-branes \cite{Witten:1996hc}. Similarly, we hope it will help understand better the contribution of five-brane instantons to effective three and four dimensional supergravities \cite{Becker:1995kb, Witten:1996bn, Dijkgraaf:2002ac, Tsimpis:2007sx, Donagi:2010pd, Alexandrov:2010ca}. Global gravitational anomalies should also put further constraints on six-dimensional supergravities containing self-dual fields \cite{Avramis:2006nb}, which should be especially relevant to the study of the landscape of six-dimensional compactifications \cite{Taylor:2011wt}. \\ 

The anomaly formula derived here has a close relative in the mathematical literature, in the work of Hopkins and Singer \cite{hopkins-2005-70}, although our derivation is completely independent of their results. In fact, even if their approach does not make any reference to the self-dual field theory, one of their motivations was to interpret and generalize a formula for the global gauge anomaly of the self-dual field coupled to an external gauge field derived by Witten in \cite{Witten:1996hc}. The generalization of Hopkins and Singer turns out to capture both the gauge and gravitational anomalies of the self-dual field theory.

Hopkins and Singer construct a certain functor between differential cohomology categories, associated to a map between topological manifolds with certain extra structure. If applied to a fiber bundle of $4\ell+2$-dimensional manifolds, this functor constructs a line bundle with connection on the base of the bundle. When the base parameterizes Riemannian metrics on the fibers, our work suggests that the inverse of the Hopkins-Singer line bundle coincides with the anomaly bundle of the self-dual field. In \cite{Monnier2011}, another characterization of the anomaly bundle was provided. Namely it was identified as the pull-back of the theta bundle living on a certain modular variety parameterizing the complex structures on the intermediate Jacobian of $M$. This bundle was endowed with a connection inherited from the determinant bundle of the Dirac operator coupled to chiral spinors. The present work therefore show a relation of the Hopkins-Singer construction to theta functions and the index theory of Dirac operators. While this relation is well-known in dimension 2, and was explicitly described in Section 2.1 of \cite{hopkins-2005-70}, it has remained mysterious in higher dimensions. We explain this a bit more in Section \ref{SecMathConj}.\\

The setup for our derivation is as follows. We consider a self-dual field theory on a manifold $M$ of dimension $4\ell+2$, with $\ell$ an integer. We endow it with a quadratic refinement of the intersection form (QRIF) on $H^{2\ell+1}(M,\mathbbm{Z})$, on which the quantum self-dual field theory is known to depend \cite{Witten:1996hc}. A QRIF can informally be thought of as a characteristic $\eta$ for a Siegel theta function on the torus $H^{2\ell+1}(M,\mathbbm{R})/H^{2\ell+1}(M,\mathbbm{Z})$. The anomaly bundle $\mathscr{A}^\eta$ is defined over $\mathcal{M}/\mathcal{D}^{(1,2)}_\eta$, the space of Riemannian metrics on $M$ quotiented by a certain subgroup of the group of diffeomorphisms on $M$. The global anomaly consists in the set of holonomies of a natural connection on this bundle.

The first step in the derivation is to obtain an anomaly formula for the fourth power of the anomaly bundle. In \cite{Monnier2011}, it was shown that $(\mathscr{A}^\eta)^2$, as a bundle with connection, is isomorphic to $\mathscr{D}^{-1} \otimes \mathscr{F}^\eta$, where $\mathscr{D}$ is the determinant bundle of the Dirac operator coupled to chiral spinors and $\mathscr{F}^\eta$ is a certain flat bundle with known holonomies. It was pointed out that this formally solves the problem of computing the global anomaly of a pair of self-dual fields, because the Bismut-Freed formalism \cite{MR853982, MR861886} allows one to compute the holonomies of $\mathscr{D}$. It turns out that the holonomy formula for $\mathscr{D}$ is very hard to use. However, $\mathscr{D}^2$ is isomorphic to the determinant bundle $\mathscr{D}_s$ of the signature operator, whose holonomy formula is much easier to handle. We have $(\mathscr{A}^\eta)^4 \simeq \mathscr{D}_s^{-1} \otimes (\mathscr{F}^\eta)^2$. Using the Bismut-Freed formula and the known holonomies of $(\mathscr{F}^\eta)^2$, we obtain a holonomy formula for $(\mathscr{A}^\eta)^4$. This formula is expressed by means of the mapping torus $\hat{M}_c$ associated with the loop $c$ in $\mathcal{M}/\mathcal{D}^{(1,2)}_\eta$ along which we are trying to compute the holonomy. Its logarithm is a sum of the eta invariant $\eta_0$ of the Atiyah-Patodi-Singer Dirac operator on $\hat{M}_c$, the dimension $h$ of its space of zero modes and a certain character of the theta group $\Gamma^{(1,2)}_\eta$ associated to the characteristic $\eta$, describing the holonomies of $(\mathscr{F}^\eta)^2$.

The second step consists in reexpressing the character in terms of topological data. It turns out that by combining it with $h$, we obtain the Arf invariant of $\hat{M}_c$. The Arf invariant is a certain topological invariant defined in terms of a quadratic refinement of the linking pairing on the torsion cohomology of $\hat{M}_c$, determined by the QRIF $\eta$. It is a generalization of the Rohlin invariant of spin manifolds of dimension $8\ell+3$ to manifolds of dimension $4\ell+3$ \cite{Lee1988}.

We then have to take a fourth root of the resulting expression. As the holonomies are $U(1)$-valued, this is a very delicate operation. The most obvious choice of fourth roots seems to be the correct one, although we can only offer consistency checks. In this way, we obtain a holonomy formula for $\mathscr{A}^\eta$, up to a subtle ambiguity that was already present in \cite{Monnier2011}, and that only reflects the lack of a complete definition of the self-dual field theory on an arbitrary compact oriented Riemannian manifold.

Our derivation is not fully rigorous. We explain the reasons why it fails to be so in Section \ref{SecDeriv}. Also, the final anomaly formula depends on a ``integral lift of the Wu class'', written $\lambda_\eta$ in the following, which is not determined explicitly by our derivation. In the case of spin manifolds of dimension 2 or 10, $\lambda_\eta$ can be taken to vanish for a preferred choice of $\eta$. In the case of spin manifolds of dimension 6, there is a proposal for what $\lambda_\eta$ should be, again for a specific $\eta$, but it is not clear to us that the resulting holonomies are compatible with the known local anomaly of the self-dual field. This is the sense in which the case of dimension 6 is still open. We explain this in more detail in Sections \ref{SecFourthRoot} and \ref{SecSpinManif}. The general construction of $\lambda_\eta$ for an arbitrary QRIF $\eta$ on non-spin manifolds remains an important open problem. Although computations can already be made with the formula we derive here, the general construction of $\lambda_\eta$ will have to be made explicit in order to gain a complete and practical knowledge of the global gravitational anomaly of the self-dual field theory. \\

The paper is organized as follows. Section \ref{SecSummAnomSDF} is a review of the current knowledge about the gravitational anomaly of the self-dual field. We review its local anomaly \cite{AlvarezGaume:1983ig}, Witten's formula for the global anomaly \cite{Witten:1985xe}, valid when $M$ has no middle-degree cohomology, and our recent work on the topological anomaly \cite{Monnier2011}. Section \ref{SecMathBack} presents some mathematical background that we need for the derivation. We review the modular geometry associated to the intermediate Jacobian of $M$ and present results on the topological Picard groups of certain modular varieties. We then move to quadratic refinements and their relations to manifolds of dimension $4\ell+2$, $4\ell+3$ and $4\ell+4$. Specializing to dimension $4\ell+3$, we review the results of \cite{Lee1988} about the Arf invariant of mapping tori. In Section \ref{SecGlobGravAnom}, we derive the anomaly formula and point out that the anomaly bundle coincides with the inverse the Hopkins-Singer bundle. The anomaly formula is presented in a self-contained way in Section \ref{SecFormGGA}. We also discuss the inclusion of gauge anomalies in the formula and the case when the self-dual field lives on a spin manifold. In Section \ref{SecDissc}, we discuss in which sense our derivation fails to be fully rigorous and we present the mathematical insight we gained as a conjecture. Then we check anomaly cancellation in type IIB supergravity and present a few other expected physical applications. An appendix describes the cohomology of mapping tori.

After a look at Section \ref{SecSummAnomSDF}, the reader uninterested in the math and in the derivation can jump to Section \ref{SecFormGGA}, where the anomaly formula is presented together with the minimal background required for its understanding. Sections \ref{SecCombGGA} and \ref{SecSpinManif} contain as well important information for physical applications, and Section \ref{SecGlobAnCan} provides a concrete example of a check of global gravitational anomaly cancellation.

\section{A review of the current knowledge}

\label{SecSummAnomSDF}

\subsection{The setup}

\label{SecSetup}

We consider a Euclidean $2\ell$-form field with self-dual $2\ell+1$-form field strength on a $4\ell+2$-dimensional oriented compact manifold $M$. 
The space of Riemannian metrics on $M$ will be denoted $\mathcal{M}$. The group of diffeomorphisms of $M$ leaving fixed both a given point on $M$ and its tangent space will be denoted by $\mathcal{D}$. Note that such diffeomorphisms necessarily preserve the orientation of $M$. With this restriction on the group of diffeomorphisms, the quotient $\mathcal{M}/\mathcal{D}$ is smooth \cite{Alvarez:1984yi}.

$\mathcal{M}/\mathcal{D}$ can be thought of as a space of parameters for a quantum field theory defined on $M$. As a result, we expect the quantum partition function of the theory to be a function over $\mathcal{M}/\mathcal{D}$. It turns out that in general, it is only the section of a certain line bundle over $\mathcal{M}/\mathcal{D}$, which we will call the anomaly bundle in the following. This line bundle comes with a natural Hermitian structure and compatible connection $\nabla$. The ``local anomaly'' coincides with the curvature of $\nabla$, while the ``global anomaly'' is given by its holonomies. If the local anomaly does not vanish, then the holonomy of $\nabla$ depends on a choice of cycle in $\mathcal{M}/\mathcal{D}$, and not only on its homotopy class.

Anomalies are usually encountered in chiral fermionic theories. The would-be partition function is formally the determinant of a chiral Dirac operator $D$ and anomalies are associated with the fact that there is no covariant way of defining such a determinant. Consider the fiber bundle $(M \times \mathcal{M})/\mathcal{D}$, where $\mathcal{D}$ acts by its defining action on $M$ and by pull-backs on $\mathcal{M}$. \footnote{Strictly speaking, in this paragraph we should choose $M$ to be spin, pick a spin structure and consider only diffeomorphisms preserving the spin structure. In the following $M$ will not necessarily be spin.} The general construction of Bismut and Freed \cite{MR853982, MR861886} applied to a family of Dirac operators $D$ on $(M \times \mathcal{M})/\mathcal{D}$ allows one to construct a determinant line bundle $\mathscr{D}$ over $\mathcal{M}/\mathcal{D}$, equipped with a natural Hermitian metric and a compatible connection $\nabla_{\mathscr{D}}$. When the index of the chiral Dirac operator is zero, there is a canonical section, which can be identified with the determinant of the chiral Dirac operator. Bismut and Freed provide explicit formulas for the curvature and holonomies of $\nabla_{\mathscr{D}}$, recovering the formulas for the local and global anomalies derived previously using physical methods in \cite{AlvarezGaume:1983ig,Witten:1985xe}. These works essentially solve the problem of the computation of gravitational anomalies for chiral fermionic theories.

However the self-dual field is not a fermionic theory. Formally, it is possible to supplement the theory of a self-dual $2\ell+1$-form with non-anomalous self-dual forms of mixed degree $(p, 4\ell+2-p)$ for $p = 0,1,..,2\ell$. If we write $\mathscr{S}_+$ and $\mathscr{S}_-$ for the even and odd components of the spinor bundle $\mathscr{S}$ of $M$ \footnote{We do not require $M$ to be spin. If it is not, the following isomorphism is still valid locally on $M$.}, the space of self-dual forms (of all degrees) is isomorphic to $\mathscr{S} \otimes \mathscr{S}_+$. This suggests that the anomaly of the self-dual field should be describable in the same formalism as the fermionic theories, using the Dirac operator coupled to chiral spinors:
\be
\label{EqDirChirSpin}
D: \mathscr{S}_+ \otimes \mathscr{S}_+ \rightarrow \mathscr{S}_- \otimes \mathscr{S}_+ \;.
\ee
We will see that this expectation is fulfilled in the case of the local anomaly, but not quite in the case of the global anomaly. From now on, $D$ and $\mathscr{D}$ will always refer to the Dirac operator coupled to chiral spinors and its determinant bundle.

\subsection{The local anomaly}

\label{SecLocAnom}

The local anomaly was computed \cite{AlvarezGaume:1983ig, AlvarezGaume:1984dr, AlvarezGaume:1983cs, Alvarez:1984yi} using the properties of the Dirac operator \eqref{EqDirChirSpin}. A small complication stems from the fact that the Dirac operator \eqref{EqDirChirSpin} is complex in Euclidean signature. The cure found was to divide the formula for the local anomaly by two to take into account the real projection that exists in Lorentzian signature. The anomaly formula then reads
\be
\label{EqCurvDetBd}
R_{\mathscr{A}} = 2\pi i \left(\int_M \frac{1}{8}L(R_{TM})\right)^{(2)} \;.
\ee 
$R_{\mathscr{A}}$ denotes the curvature of the anomaly bundle $\mathscr{A}$ for a self-dual field, equal to minus half the curvature of the natural connection $\nabla_{\mathscr{D}}$ on $\mathscr{D}$. The sign flip between the curvature of $\mathscr{A}$ and $\mathscr{D}$ is due to the fact that the partition function of a pair of self-dual fields is the inverse of the partition function of the fermionic theory whose anomaly bundle is $\mathscr{D}$. $R_{TM}$ is the curvature of $TM$, seen as a bundle over $(M \times \mathcal{M})/\mathcal{D}$. The exponent $(\bullet)^{(2)}$ denotes the projection on the two-form component and the L-genus is defined by
\be
\label{EqL-Genus}
L(R) = 2^{2\ell+2}{\rm det}^{1/2} \frac{R/4\pi}{\tanh{R}/4\pi} \;.
\ee 

Later, the local gravitational anomaly was also derived in \cite{PhysRevLett.63.728} from the Henneaux-Teitelboim action \cite{Henneaux:1988gg} for the self-dual field, and in \cite{Monnier2011} from an action closely related to the Belov-Moore action \cite{Belov:2006jd}. In \cite{Monnier:2010ww}, the local anomaly was recovered in the framework of geometric quantization. Moreover, the norm of the partition function of the self-dual field was shown to coincide with the square root of the norm of the determinant section of $\mathscr{D}^{-1}$, indicating that the Hermitian structures of $\mathscr{A}^2$ and $\mathscr{D}^{-1}$ coincide.

The major insight that can be gained from these works is that the curvature of the connection $\nabla_\mathscr{A}$ on $\mathscr{A}$ coincides with minus half the curvature of $\nabla_{\mathscr{D}}$.

\subsection{Witten's formula for the global anomaly}

\label{SecWittFormGlobAn}

The global anomaly is more subtle to understand, because it is sensitive to torsion. Partial results have been obtained by Witten in \cite{Witten:1985xe}, in the case when the cohomology of $M$ in degree $2\ell+1$ vanishes. In this case, $\mathscr{A}^2$ is topologically trivial and coincides with $\mathscr{D}^{-1}$ \cite{Monnier2011}. Also, the complications associated with the zero modes of the self-dual field disappear. Instead of repeating Witten's arguments, we reframe them in the formalism of Bismut-Freed \cite{MR861886} and spell out the subtleties which occur. 

In principle, we can use the Bismut-Freed formula in order to compute the holonomies of $\nabla_{\mathscr{D}}$. However, this turns out to be of little practical interest. Instead, the good strategy is to compute the holonomy of the Bismut-Freed connection $\nabla_{\mathscr{D}_s}$ associated to the signature operator $D_s$:
\be
\label{EqDirSign}
D_s: \mathscr{S}_+ \otimes \mathscr{S} \rightarrow \mathscr{S}_- \otimes \mathscr{S} \;.
\ee
We denote the determinant bundle of $D_s$ by $\mathscr{D}_s$. Recall that self-dual forms on $M$ can be identified with $\mathscr{S}_+ \otimes \mathscr{S}_+ \oplus \mathscr{S}_- \otimes \mathscr{S}_+$. As complex conjugation turns self-dual forms on $M$ into anti self-dual forms, we see that the determinant bundle $\mathscr{D}_s$ of $D_s$ can be expressed as
\be
\mathscr{D}_s = \mathscr{D} \otimes \bar{\mathscr{D}}^\dagger \simeq \mathscr{D}^2 \;,
\ee 
where the bar denotes complex conjugation. Hence the holonomies of $\nabla_{\mathscr{D}_s}$ are the squares of those of $\nabla_{\mathscr{D}}$.

Consider a loop $c$ in $\mathcal{M}/\mathcal{D}$, along which we would like to compute the holonomy of the Bismut-Freed connection $\nabla_{\mathscr{D}_s}$. Such a loop is naturally associated to a diffeomorphism $\phi_c$ of $M$. Indeed, it lifts to an open path in $\mathcal{M}$, with the two endpoints differing by the action of an element of $\mathcal{D}$. The restriction of the fiber bundle $(M \times \mathcal{M})/\mathcal{D}$ to $c$ defines a $4\ell +3$-dimensional manifold $\hat{M}_c$. This manifold can be pictured as $M \times [0,1]$ quotiented by the relation $(x,0) \simeq (\phi_c(x),1)$ and is called the \emph{mapping torus} of $\phi_c$. It is naturally endowed with a metric $g_M$ on its fibers. Choose an arbitrary metric $g_\circ$ on $S^1$, and rescale it to $g_\circ/\epsilon^2$, $\epsilon > 0$. $g_\epsilon := g_\circ/\epsilon^2 \oplus g_M$ is a family of metrics on $\hat{M}_c$. 

Denote by $\hat{\mathscr{S}}$ the spinor bundle of $\hat{M}_c$ and consider the Dirac operator on $\hat{M}_c$ twisted by $\hat{\mathscr{S}}$ \footnote{This is \emph{not} the Dirac operator defined in \cite{MR861886}, which should in principle enter the Bismut-Freed formula. Indeed, using the latter would lead to impractical formulas for checking anomaly cancellation. It is expected that the holonomies $\exp \pi i (\eta + h)$ for the two operators agree in the limit $\epsilon \rightarrow 0$, but we have not been able to prove this. More details can be found in Section \ref{SecDeriv}. We thank Dan Freed for a correspondence about this point.}
\be
\label{Eq4l3DO}
\hat{D}_\epsilon: \hat{\mathscr{S}} \otimes \hat{\mathscr{S}} \rightarrow \hat{\mathscr{S}} \otimes \hat{\mathscr{S}} \;.
\ee
$\hat{D}_\epsilon$ depends on $\epsilon$ because the Levi-Civita connection of $g_\epsilon$ enters its definition. The eta invariant $\eta_\epsilon$ of $\hat{D}_\epsilon$ is defined as the value of the analytic continuation of
\be
\sum_{\lambda \in {\rm Spec} (\hat{D}_\epsilon)} \frac{{\rm sgn}(\lambda)}{|\lambda|^s}
\ee
at $s = 0$ \cite{MR0397797}. The holonomy formula for the Bismut-Freed connection on the determinant bundle of the signature operator reads \cite{MR861886, Freed:1986zx}
\be
\label{EqHolNZModes}
{\rm hol}_{\mathscr{D}_s}(c) = (-1)^{{\rm index} D_s} \lim_{\epsilon \rightarrow 0} \exp -\pi i (\eta_\epsilon + h_\epsilon) \;,
\ee
where $h_\epsilon$ is the dimension of the kernel of $\hat{D}_\epsilon$. As $D_s$ maps self-dual forms to anti self-dual forms, its index modulo 2 necessarily vanishes and the first factor does not contribute. We need a more explicit formula for $\hat{D}_\epsilon$ \cite{MR0397797}:
\be
\label{EqExplFormDHat}
\hat{D}_\epsilon = (-1)^{k+\ell+1}\big(d \ast_\epsilon - \ast_\epsilon d \big)|_{\Omega^{\rm even}(\hat{M}_c,\mathbbm{R})} \;.
\ee
We can model the bundle $\hat{\mathscr{S}} \otimes \hat{\mathscr{S}}$ with $\Omega^{\rm even}(\hat{M}_c,\mathbbm{R})$, and it turns out that in this model $\hat{D}_\epsilon$ takes the form \eqref{EqExplFormDHat}, where $\ast_\epsilon$ is the $4\ell+3$-dimensional Hodge star operator associated to the metric $g_\epsilon$. $k$ is equal to $p/2$, where $p$ is the degree of the form acted on by $\hat{D}_\epsilon$. This expression for $\hat{D}_\epsilon$ is extremely useful to compute $h_\epsilon$. Indeed, the kernel of $\hat{D}_\epsilon$ is simply the space of harmonic even forms, which we can identify with the cohomology $H^{\rm even}(\hat{M}_c,\mathbbm{R})$. In particular, its dimension is independent of the metric and of $\epsilon$. We compute the cohomology groups of $\hat{M}_c$ in terms of the cohomology groups of $M$ and the action of $\phi_c$ in appendix \ref{SecCohomMapTor}. When there is no cohomology in degree $2\ell+1$, $h_\epsilon$ is even. The Bismut-Freed formula therefore reads in this case
\be
\label{EqHolNZModes2}
{\rm hol}_{\mathscr{D}_s}(c) = \lim_{\epsilon \rightarrow 0} \exp -\pi i \eta_\epsilon \;.
\ee
Although this is not the case for the eta invariant of a generic Dirac operator, $\eta_\epsilon$ has a well-defined limit when $\epsilon \rightarrow 0$, which we write $\eta_0$ from now on.

Now recall that as bundles with connection, $\mathscr{D}_s \simeq \mathscr{D}^2$. Moreover, $\mathscr{A}^2 \simeq \mathscr{D}^{-1}$ in case the cohomology of degree $2\ell+1$ of $M$ vanishes. So in order to obtain a formula for the holonomy of the connection on $\mathscr{A}$, we have to take a fourth root of \eqref{EqHolNZModes2}. As the holonomies are valued in $U(1)$, this is a non-trivial operation to perform. For each homotopy class of loops, we have a priori four equally valid choices of fourth roots. Witten's formula amounts to the following simple set of choices of fourth roots, which we cannot motivate at this point:
\be
\label{EqHolNZModes3}
{\rm hol}_{\mathscr{A}}(c) = \exp \frac{\pi i}{4} \eta_0 \;.
\ee
This formula is still impractical for checking anomaly cancellations. Indeed, computing explicitly the eta invariant is a hopeless task.

However, we can do better if the mapping torus $\hat{M}_c$ is the boundary of a $4\ell+4$-dimensional Riemannian manifold $W$. In this case, the Atiyah-Patodi-Singer theorem \cite{MR0397797} provides an alternative expression for the eta invariant:
\be
\label{EqAPS}
\eta_\epsilon = \int_W L(R_{W,\epsilon}) - \sigma_W \;.
\ee
In this formula, we have endowed $W$ with a metric that restricts to $g_\epsilon$ on the boundary $\hat{M}_c$, and tends to a product near $\hat{M}_c$. $R_{W,\epsilon}$ is the corresponding curvature tensor. The $L$ genus takes the form \eqref{EqL-Genus} and $\sigma_W$ is the signature of the symmetric intersection form on the relative cohomology $H^{2\ell+2}(W,\hat{M}_c,\mathbbm{Z})$. Writing $L_0$ for $\lim_{\epsilon \rightarrow 0} L(R_{W,\epsilon})$, we get the final holonomy formula \cite{Witten:1985xe}
\be
\label{EqHolNZModes4}
{\rm hol}_{\mathscr{A}}(c) = \exp \frac{2\pi i}{8} \left(\int_W L_0  - \sigma_W\right) \;.
\ee

The logarithm of this formula takes the form of the sum of a topological invariant of $W$ and the integral of the local density appearing in the local anomaly formula \eqref{EqCurvDetBd}. Such a form is very practical for checking the cancellation of global gravitational anomalies. Indeed, suppose that we have a theory combining self-dual fields with chiral fermions in such a way that the local anomaly vanishes. We can compute the global gravitational anomaly for this theory by adding the logarithms of the equivalents of equation \eqref{EqHolNZModes4} for each of the component theories. As the local anomaly vanishes, we know that the terms containing index densities will cancel, and we are simply left with a sum of topological invariants of the manifold $W$. The global anomaly vanishes if and only if the sum of these topological invariants is an integer multiple of $2\pi i$ for all possible pairs $(\hat{M}_c, W)$. Investigating the integrality of the sum of topological invariant is a much simpler problem than computing the eta invariants explicitly. This convenience has a price: we can compute the holonomies only on cycles $c$ such that the corresponding mapping torus $\hat{M}_c$ bounds a manifold $W$. Note that the case when $\hat{M}_c$ is a non-zero torsion element of the bordism group (i.e. a multiple of the corresponding cycle is a boundary) can be treated using the techniques of \cite{Freed:1986zx, MR936082}.

We can also see why it would have been a bad idea to apply the Bismut-Freed formula to $D$, the Dirac operator coupled to chiral spinors. First, we would not have been able to compute easily the dimension of its kernel, which is not fixed by cohomology considerations as in the case of $D_s$. Second, we would not have been able to use the Atiyah-Patodi-Singer theorem in any obvious way in order to reexpress the eta invariant of $D$ in terms of manageable quantities. 

The formula \eqref{EqHolNZModes4} solves the problem of computing  the global gravitational anomaly of the self-dual field on a manifold that has no cohomology in degree $2\ell+1$. Note however that there is a caveat in the argument: we made an apparently arbitrary choice of fourth root to pass from the holonomy formula of $\mathscr{D}_s$ to the one for $\mathscr{A}$. A priori, nothing guarantees that this is the right choice.

\subsection{The topological anomaly}

\label{SecHolFormPSD}

In \cite{Monnier2011}, we managed to characterize the topology of the anomaly bundle. In particular, we showed that when there is non-trivial cohomology in degree $2\ell+1$, $\mathscr{A}^2$ is no longer isomorphic to $\mathscr{D}^{-1}$. Rather, we have 
\be
\label{EqTopAnom}
(\mathscr{A^\eta})^2 \simeq \mathscr{D}^{-1} \otimes \mathscr{F}^\eta \;,
\ee
where $\mathscr{F}^\eta$ is a non-trivial flat bundle. $\eta$ is a half-integral element of $H^{2\ell+1}(M,\mathbbm{R})$, defining a quadratic refinement of the intersection form (QRIF), about which we will say more in Section \ref{SecQuadrRef}. \footnote{The QRIF $\eta$ should not to be confused with the eta invariant $\eta_\epsilon$.} For now we can simply think about it as a parameter needed to characterize the theory on manifolds with $H^{2\ell+1}(M,\mathbbm{R}) \neq 0$. \eqref{EqTopAnom} shows that Witten's formula will necessarily be modified in this case.

We deduce from \eqref{EqTopAnom} that
\be
{\rm hol}_{(\mathscr{A}^\eta)^2}(c) = \chi^\eta \, {\rm hol}_{\mathscr{D}^{-1}}(c) \quad {\rm and} \quad {\rm hol}_{(\mathscr{A}^\eta)^4}(c) = (\chi^\eta)^2 \, {\rm hol}_{(\mathscr{D}_s)^{-1}}(c) \;,
\ee
where $\chi^\eta$ is a certain character of the mapping class group of $M$ describing the holonomies of $\mathscr{F}^\eta$, which will be written explicitly in Section \ref{SecTopPicGr}. $\mathscr{F}^\eta$, and therefore $\chi^\eta$, has order 4, so we have
\be
(\mathscr{A^\eta})^8 \simeq \mathscr{D}_s^{-2} \;.
\ee
For $H^{2\ell+1}(M,\mathbbm{R})$ arbitrary, the holonomy formula for $\mathscr{D}_s^{-1}$ takes in general the form
\be
\label{EqHolDs2}
{\rm hol}_{(\mathscr{D}_s)^{-1}}(c) = \exp \pi i (\eta_0 + h) \;.
\ee
$h$ is the dimension of the kernel of $\hat{D}_s$, which does not vanish modulo 2 in general. However, it is equal to the dimension of the $\phi_c$-invariant subspace of harmonic $2\ell+1$-forms on $M$, which we will write $H^{2\ell+1}_{\phi_c}(M,\mathbbm{R})$ (see appendix \ref{SecCohomMapTor}). It is clearly independent of the metric $g_\epsilon$ on $\hat{M}_c$.

The holonomy of the anomaly bundle of the self-dual field is a certain fourth root of \eqref{EqHolDs2} twisted by the character $(\chi^\eta)^2$. In order to determine it, we need more mathematical background.

\section{Mathematical background}

\label{SecMathBack}

In this section we review various mathematical topics\footnote{Part of the relevant material appeared in the review \cite{Sati:2011pg}.} that will prove necessary in the derivation of the anomaly formula. Part of the material in Sections \ref{SecIntJacModG} and \ref{SecTopPicGr} is available in a slightly expanded form in Section 4 of \cite{Monnier2011}.

\subsection{The intermediate Jacobian and modular geometry}

\label{SecIntJacModG}

Recall that the self-dual field theory is defined on an oriented compact Riemannian manifold $M$, of dimension $4\ell+2$. In this dimension, the intersection product defines a symplectic form on $\Omega^{2\ell+1}(M,\mathbbm{R})$:
\be
\omega(R,T) = 2\pi \int_M R \wedge T \;.
\ee
The Hodge star operator constructed from the Riemannian metric squares to $-1$ on $\Omega^{2\ell+1}(M,\mathbbm{R})$, so it defines a complex structure, which is compatible with the symplectic form. $\Omega^{2\ell+1}(M,\mathbbm{R})$ is therefore Kähler. These structures pass to the subspace of harmonic forms, hence to the cohomology $H^{2\ell+1}(M,\mathbbm{R})$ by Hodge's theorem. We will write $2n$ for the dimension of $H^{2\ell+1}(M,\mathbbm{R})$. The free quotient of $H^{2\ell+1}(M,\mathbbm{Z})$, which we will write $\Lambda$, is an integral lattice in $H^{2\ell+1}(M,\mathbbm{R})$. It is possible to choose a Darboux basis $\{\alpha_i,\beta^i\} \subset H^{2\ell+1}(M,\mathbbm{R})$ satisfying 
\be
\omega(\alpha_i, \alpha_j) = \omega(\beta^i,\beta^j) = 0 \;, \qquad \omega(\alpha_i,\beta^j) = 2 \pi \delta_i^j \;.
\ee
The lattices generated by $\{\alpha_i\}$ and $\{\beta^i\}$ will be denoted by $\Lambda_1$ and $\Lambda_2$. We introduce coordinates $a^i$, $b_i$ such that 
\be
\alpha_i = \frac{\partial}{\partial a^i} \;, \qquad \beta^i = \frac{\partial}{\partial b_i} \;.
\ee
We will write $(v_1, v^2)$ for the components of $v \in H^{2\ell+1}(M,\mathbbm{R})$ on the Lagrangian subspaces containing $\Lambda_1$ and $\Lambda_2$, respectively.

The Lazzeri intermediate Jacobian $\mathcal{J}$ of $M$ \cite{MR1785408, MR1713785} is defined to be the torus
\be
\mathcal{J} := H^{2\ell+1}(M,\mathbbm{R})/\Lambda \;,
\ee
endowed with the complex structure defined by the Hodge star operator. Because of the positive definiteness of the bilinear form $\omega(\,.\,,\,\ast.\,)$, $\mathcal{J}$ is an Abelian variety, in fact a principally polarized one. In case $M$ is Kähler and does not admit any non-primitive cohomology in degree $2\ell+1$, $\mathcal{J}$ coincides with the Weil intermediate Jacobian.

Associated with the abelian variety $\mathcal{J}$, there is a moduli space $\mathcal{C}$ parameterizing the compatible complex structures of $\mathcal{J}$ modulo homeomorphisms homotopic to the identity. $\mathcal{C}$ is a Siegel upper half-space; it can be pictured as the space of complex $n$ by $n$ symmetric matrices $\tau$ with positive definite imaginary part. $\tau$ is defined such that the holomorphic coordinates $z^i$ on $H^{2\ell+1}(M,\mathbbm{R})$ in the corresponding complex structure read
\be
\label{EqDefz}
z_i = \tau_{ij}a^i + b_i \;.
\ee
$\mathcal{C}$ is a contractible space.

The group of diffeomorphisms $\mathcal{D}$ of $M$ acts on $H^{2\ell+1}(M,\mathbbm{R})$, and therefore on $\mathcal{J}$, through its action by pull-back on harmonic forms. This action maps $\Lambda$ to itself and preserves $\omega$. It therefore factorizes through the action of ${\rm Sp}(2n,\mathbbm{Z})$ on $H^{2\ell+1}(M,\mathbbm{R})$. The connected component of the identity $\mathcal{D}_0$ acts trivially, so the action factorizes as well through the mapping class group $\mathcal{D}/\mathcal{D}_0$ of $M$:
\be
\label{EqProjMCGSp}
\mathcal{D} \rightarrow \mathcal{D}/\mathcal{D}_0 \rightarrow {\rm Sp}(2n,\mathbbm{Z}) \hookrightarrow H^{2\ell+1}(M,\mathbbm{R}) \;,
\ee
where the hook arrow denotes the action. We can write 
\be
\label{EqDecBlockGamma}
\gamma = \begin{pmatrix} A \!& B \\ C \!& D \end{pmatrix} \;,
\ee
for the matrix of an element $\gamma \in {\rm Sp}(2n,\mathbbm{Z})$, where the block decomposition is induced by the decomposition $\Lambda = \Lambda_1 \oplus \Lambda_2$. The blocks satisfy $A^tB = B^tA$, $C^tD = D^tC$ and $AD^t - BC^t = \mathbbm{1}_n$. The action on $H^{2\ell+1}(M,\mathbbm{R})$ then reads explicitly
\be
\gamma \cdot \alpha_i = A_i^{\;j} \alpha_j + B_{ij} \beta^j \;, \quad \gamma \cdot \beta^i = C^{ij} \alpha_j + D^i_{\;j} \beta^j \;.
\ee
The action on $H^{2\ell+1}(M,\mathbbm{R})$ induces an action on $\mathcal{C}$, given by
\be
\label{EqActHP}
\qquad \gamma.\tau = (A\tau + B)(C\tau + D)^{-1} \;. 
\ee

$\mathcal{J}$ also inherits the Kähler structure of $H^{2\ell+1}(M,\mathbbm{R})$. We will still write $\omega$ for the corresponding Kähler 2-form. The holomorphic line bundles over $\mathcal{J}$ whose first Chern classes coincide with $\omega$ are parameterized by characteristics $\eta \in H^{2\ell+1}(M,\mathbbm{R})$, and any two bundles whose characteristics differ by an element of $\Lambda$ are isomorphic \cite{MR2062673}. We write these bundles $\mathscr{L}^\eta$. $\mathscr{L}^\eta$ has a unique section up to scalar multiples, whose pull-back to $H^{2\ell+1}(M,\mathbbm{R})$ is given by the Siegel theta function 
\be
\label{EqDefThetaFunc}
\theta^\eta(z, \tau) = \sum_{r \in \Lambda_1 + \eta_1} \exp \big(\pi i r^i \tau_{ij} r^j + 2\pi i (z_i + \eta^2_i)r^i \big) \;,
\ee
$z \in H^{2\ell+1}(M,\mathbbm{R})$. 

The ${\rm Sp}(2n,\mathbbm{Z})$ action on $H^{2\ell+1}(M,\mathbbm{R})$ induces an action on the line bundles $\mathscr{L}^\eta$ that permutes bundles with different characteristics. The resulting action on the characteristics is affine:
\be
\label{EqTransEta}
\gamma \ast \eta = (\gamma^t)^{-1} \cdot \eta + \frac{1}{2}((C D^t)_0)^i \alpha_i + \frac{1}{2}((A B^t)_0)_i \beta^i \;,  
\ee
where the notation $(M)_0$ denotes the vector formed by the diagonal entries of the matrix $M$. From now on we will restrict ourselves to half-integral characteristics, for reasons that will be apparent in Section \ref{SecQuadrRef}. The action \eqref{EqTransEta} on half-integral characteristics decomposes into two orbits. The characteristics in the orbit of $\eta = 0$ are called even, while the characteristics in the other orbit are called odd. We see immediately that for an element of ${\rm Sp}(2n,\mathbbm{Z})$ to preserve $\mathscr{L}^{\eta = 0}$, we need $(AB^t)_0 = (CD^t)_0 = 0 \; {\rm mod} \; 2$. The group of such symplectic transformations is known as the theta group and will be denoted by $\Gamma^{(1,2)}$. In order for an element of ${\rm Sp}(2n,\mathbbm{Z})$ to preserve all the bundles $\mathscr{L}^{\eta}$, we need the stronger condition $\gamma = \mathbbm{1}_{2n}$ modulo 2. The corresponding group is known as the level 2 congruence subgroup of ${\rm Sp}(2n,\mathbbm{Z})$, and will be written $\Gamma^{(2)}$. We will write $\mathcal{D}^{(1,2)}$ and $\mathcal{D}^{(2)}$ for the subgroups of $\mathcal{D}$ that are mapped according to \eqref{EqProjMCGSp} into $\Gamma^{(1,2)}$ and $\Gamma^{(2)}$, respectively. Given any half-integral characteristic $\eta$, there is a maximal subgroup of ${\rm Sp}(2n,\mathbbm{Z})$ leaving it fixed, which we write $\Gamma_\eta^{(1,2)}$. We write $\mathcal{D}^{(1,2)}_\eta$ for its preimage in $\mathcal{D}$ under \eqref{EqProjMCGSp}.

\subsection{The topological Picard groups of certain modular varieties}

\label{SecTopPicGr}

Let us now consider the quotients $\mathcal{T}^{(2,1)} := \mathcal{C}/\Gamma^{(2,1)}$ and $\mathcal{T}^{(2)} := \mathcal{C}/\Gamma^{(2)}$. These quotients are smooth except for orbifold singularities. Their fundamental group is given by $\Gamma^{(2,1)}$ and $\Gamma^{(2)}$, because $\mathcal{C}$ is contractible. As $\Gamma^{(2)} \subset \Gamma^{(1,2)}$ with a finite index, $\mathcal{T}^{(2)}$ is a finite covering of $\mathcal{T}^{(2,1)}$. Similarly, we define $\mathcal{T}_\eta^{(2,1)}:= \mathcal{C}/\Gamma^{(2,1)}_\eta$.

Given a finite index subgroup $\Gamma$ of ${\rm Sp}(2n,\mathbbm{Z})$ and the associated quotient $\mathcal{T} := \mathcal{C}/\Gamma$, we define a line bundle over $\mathcal{T}$ to be a $\Gamma$-equivariant line bundle over $\mathcal{C}$. This definition has the advantage of being insensitive to the possible orbifold singularities $\mathcal{T}$ may have. Under the tensor product, the set of all line bundles over $\mathcal{T}$ forms an abelian group, the topological Picard group ${\rm Pic}(\mathcal{T})$. \footnote{The usual Picard group classifies holomorphic line bundles. In the context of algebraic geometry, the topological Picard group is called the Néron-Severi group.} Our aim in this section is to present results about the topological Picard groups of $\mathcal{T}^{(2)}$ and $\mathcal{T}^{(2,1)}$. More details can be found in Section 4 of \cite{Monnier2011}.

In \cite{2009arXiv0908.0555P}, it was shown that provided $n \geq 3$, ${\rm Pic}(\mathcal{T}^{(2)})$ fits in the following short exact sequence:
\be
\label{EqSeqPic}
0 \rightarrow {\rm Hom}\big(\Gamma_{\rm ab}^{(2)}, \mathbbm{Q}/\mathbbm{Z}\big) \rightarrow {\rm Pic}(\mathcal{T}^{(2)}) \rightarrow \mathbbm{Z} \rightarrow 0 \;,
\ee
where $\Gamma_{\rm ab}^{(2)}$ denotes the abelianization of $\Gamma^{(2)}$. $\Gamma_{\rm ab}^{(2)}$ was computed in the Section 2 of \cite{Sato01012010}: 
\be
\Gamma_{\rm ab}^{(2)} = (\mathbbm{Z}_2)^{2n^2-n} \times (\mathbbm{Z}_4)^{2n} \;. 
\ee
In the exact sequence above, ${\rm Hom}\big(\Gamma_{\rm ab}^{(2)}, \mathbbm{Q}/\mathbbm{Z}\big)$ classifies flat line bundles on $\mathcal{T}^{(2)}$. The projection of $\Gamma^{(2)}$ on its abelianization can be described explicitly. Let us decompose an element $\gamma \in \Gamma^{(2)}$ into $n \times n$ blocks as follows:
\be
\label{EqDecBlockGamma2}
\gamma - \mathbbm{1} = 2 \begin{pmatrix} \tilde{A}(\gamma) \!& \tilde{B}(\gamma) \\ \tilde{C}(\gamma) \!& \tilde{D}(\gamma) \end{pmatrix} \;.
\ee
There is a homomorphism $m : \Gamma^{(2)} \rightarrow (\mathbbm{Z}_2)^{2n^2-n} \times (\mathbbm{Z}_4)^{2n}$ given by
\begin{align}
\label{EquMapAb}
m(\gamma) =  \big( &\, \{\tilde{A}_{ij}(\gamma)\}_{1 \leq i \leq n, 1 \leq j \leq n} \; {\rm mod} \; 2, \notag \\
&\, \{\tilde{B}_{ij}(\gamma)\}_{1 \leq i < j \leq n} \; {\rm mod} \; 2, \notag \\
&\,\{\tilde{C}_{ij}(\gamma)\}_{1 \leq i < j \leq n} \; {\rm mod} \; 2,  \\
&\,\{\tilde{B}_{ii}(\gamma)\}_{1 \leq i \leq n} \; {\rm mod} \; 4, \notag \\
&\,\{\tilde{C}_{ii}(\gamma)\}_{1 \leq i \leq n} \; {\rm mod} \; 4 \big) \notag \;.
\end{align}
The components of the map $m$ form a basis for the (additive) characters of $\Gamma^{(2)}$, or equivalently for the group of flat line bundles on $\mathcal{T}^{(2)}$. These components will be referred to as the elementary characters of $\Gamma^{(2)}$. We also define the following system of generators of $\Gamma^{(2)}$ \cite{Igusa1964}, which we will call the elementary generators:
\begin{itemize}
	\item $\alpha^{(ij)}$, the $2n \times 2n$ identity matrix with the entry $(i,j)$ replaced by $2$ and the entry $(n+j, n+i)$ replaced by $-2$, for $1 \leq i \leq n$, $1 \leq j \leq n$ and $i \neq j$;
	\item $\alpha^{(ii)}$, the $2n \times 2n$ identity matrix with the entries $(i,i)$ and $(n+i, n+i)$ replaced by $-1$, for $1 \leq i \leq n$;
	\item $\beta^{(ij)}$, the $2n \times 2n$ identity matrix with the entries $(i,n+j)$ and $(j, n+i)$ replaced by $2$ , for $1 \leq i \leq n$, $1 \leq j \leq n$ and $i < j$;
	\item $\gamma^{(ij)} := (\beta^{(ij)})^t$;
	\item $\beta^{(ii)}$, the $2n \times 2n$ identity matrix with the entry $(i,n+i)$ replaced by $2$, for $1 \leq i \leq n$;
	\item $\gamma^{(ii)} := (\beta^{(ii)})^t$.
\end{itemize}
For each elementary additive character forming the components of the map $m$, there is one of the elementary generators that is mapped to $1$ by this character and to $0$ by all the other characters. Given any character of $\Gamma^{(2)}$, its evaluation on the elementary generators allows one to express it in terms of the elementary additive characters.

The abelianization of the theta group, or equivalently the group of flat line bundles over $\mathcal{T}^{(1,2)}$, was computed in \cite{springerlink:10.1007/BF01231183}, again for $n \geq 3$:
\be
\Gamma^{(1,2)}_{\rm ab} = \mathbbm{Z}_4 \;.
\ee
This group is generated by the conjugacy class of an anisotropic transvection (see Section 4 of \cite{Monnier2011}). As $\Gamma^{(2)} \subset \Gamma^{(1,2)}$, we have an inclusion of character groups 
\be
{\rm Hom}\big(\Gamma_{\rm ab}^{(1,2)}, \mathbbm{Q}/\mathbbm{Z}\big) \rightarrow {\rm Hom}\big(\Gamma_{\rm ab}^{(2)}, \mathbbm{Q}/\mathbbm{Z}\big) \;,
\ee
corresponding to pulling back flat line bundles from $\mathcal{T}^{(1,2)}$ to $\mathcal{T}^{(2)}$. The image of the generator $\chi^0$ of ${\rm Hom}\big(\Gamma_{\rm ab}^{(1,2)}, \mathbbm{Q}/\mathbbm{Z}\big)$, corresponding to a flat bundle $\mathscr{F}^0$, is given by
\be
\label{EqCharKappa0PB}
\chi^0(\gamma) = \exp\left(\pi i \sum_{i} \tilde{A}_{ii}(\gamma)\right) \in  {\rm Hom}\big(\Gamma_{\rm ab}^{(2)}, \mathbbm{Q}/\mathbbm{Z}\big) \;.
\ee
Note that while $\mathscr{F}^0$ has order 4, it only has order 2 after it has been pulled back to $\mathcal{T}^{(2)}$. \\

Let us now consider some concrete examples of line bundles over $\mathcal{T}^{(2)}$ and $\mathcal{T}^{(1,2)}$. Define the theta constant $\theta^\eta(\tau) := \theta^\eta(z=0,\tau)$. The theta constant is the pull-back to $\mathcal{C}$ of the section of a bundle $\mathscr{C}^\eta$ on $\mathcal{T}^{(2)}$. \footnote{Theta constants vanish for odd characteristics. However their transformation formulas still define bundles $\mathscr{C}^\eta$.} Each of the bundles $\mathscr{C}^\eta$ is actually a pull-back to $\mathcal{T}^{(2)}$ of a bundle over $\mathcal{T}^{(1,2)}_\eta$ (which we still call $\mathscr{C}^\eta$). However, $\mathscr{C}^\eta$ is not the pull back from a bundle over $\mathcal{T}^{(1,2)}_{\eta'}$ if $\eta \neq \eta'$. It can be shown that the bundles $\mathscr{C}^\eta$ map to $1 \in \mathbbm{Z}$ in the short exact sequence \eqref{EqSeqPic}, so any such bundle generates the Picard group of $\mathcal{T}^{(2)}$ modulo torsion.

The Hodge bundle is defined to be the holomorphic tangent space of $\mathcal{J}$, when the latter is seen as the fiber of the universal abelian variety over $\mathcal{C}/{\rm Sp}(2n,\mathbbm{Z})$. We denote its determinant bundle by $\mathscr{K}$. As it is defined on $\mathcal{C}/{\rm Sp}(2n,\mathbbm{Z})$, $\mathscr{K}$ pulls back to $\mathcal{C}/\Gamma$ for any finite index subgroup $\Gamma$ of ${\rm Sp}(2n,\mathbbm{Z})$. In particular, it pulls back to both $\mathcal{T}^{(2)}$ and $\mathcal{T}^{(1,2)}_\eta$. Again, we will slightly abuse notation and write $\mathscr{K}$ for the pull-backs as well. $\mathscr{K}$ maps on $2 \in \mathbbm{Z}$ in \eqref{EqSeqPic}, therefore 
\be
\mathscr{F}^\eta := (\mathscr{C}^\eta)^2 \otimes \mathscr{K}^{-1}
\ee
is a flat bundle for all $\eta$. As our notation above suggests, $\mathscr{F}^{\eta=0}$ generates the group of flat line bundles over $\mathcal{T}^{(1,2)}$. Over $\mathcal{T}^{(2)}$, we can express the holonomies of the bundles $\mathscr{F}^\eta$ by means of a character $\chi^\eta$ of $\Gamma^{(2)}$. In the basis of elementary characters \eqref{EquMapAb}, the latter reads \cite{Igusa1964}
\be
\label{EqCharCeK}
\chi^\eta(\gamma) = \exp \left( \pi i \sum_j \tilde{A}_{jj}(\gamma) - 4\pi i \sum_{i} \tilde{B}_{ii}(\gamma) (\eta_1^i)^2 - 4\pi i\sum_{i} \tilde{C}_{ii}(\gamma) (\eta^2_i)^2 \right) \;.
\ee
We see that the character associated with $\mathscr{F}^0$ coincides with \eqref{EqCharKappa0PB}. The bundle $\mathscr{F}^\eta$ appearing in \eqref{EqTopAnom} is obtained by pulling back the bundle defined above from $\mathcal{T}^{(1,2)}_\eta$ to $\mathcal{M}/\mathcal{D}^{(1,2)}_\eta$. In the following, we will abuse notation and denote both bundles by $\mathscr{F}^\eta$. We will also call $\mathscr{F}^\eta$ the corresponding pull-backs on $\mathcal{T}^{(2)}$ and $\mathcal{M}/\mathcal{D}^{(2)}$.

Another class of flat bundles on $\mathcal{T}^{(2)}$ is given by $\mathscr{F}^{(\eta, \eta')} := \mathscr{C}^\eta \otimes (\mathscr{C}^{\eta'})^{-1}$. As $\mathscr{F}^{(\eta, \eta')} \simeq \mathscr{F}^{(\eta, 0)} \otimes (\mathscr{F}^{(\eta', 0)})^{-1}$, it is sufficient to study the set $\{\mathscr{F}^{(\eta, 0)}\}$. The holonomies of these bundles are easily computed from the theta transformation formulas (see for instance \cite{Monnier2011}, Section 4.5). We write the corresponding characters $\chi^{(\eta, 0)}$ and we have
\be
\label{EqCharCetaCzeroExpl}
\chi^{(\eta, 0)}(\gamma) = \exp \left( \pi i n_1(\gamma,\eta) + \frac{\pi i}{2} n_2(\gamma,\eta) \right) \;,
\ee
with
\begin{align}
n_1(\gamma,\eta)/4 = &\, \sum_{i,j} \tilde{A}_{ij}(\gamma) \eta_1^i \eta^2_j - \sum_{i < j}  \tilde{B}_{ij}(\gamma) \eta_1^i \eta_1^j - \sum_{i < j}  \tilde{C}_{ij}(\gamma) \eta^2_i \eta^2_j \;, \\
n_2(\gamma,\eta)/4 = &\, \sum_{i} \tilde{B}_{ii}(\gamma) (\eta_1^i)^2 - \sum_{i} \tilde{C}_{ii}(\gamma) (\eta^2_i)^2 \notag \;.
\end{align}

Let us pull-back $\mathscr{K}$ to $\mathcal{M}/\mathcal{D}^{(1,2)}_\eta$ and still write it $\mathscr{K}$. It is interesting to note that as topological bundles, $\mathscr{K} \simeq \mathscr{D}^{-1}$ \cite{Monnier:2010ww}. From the existence of the bundles $\mathscr{C}^\eta$, we know that $\mathscr{K}$, and hence $\mathscr{D}$, admits a square root modulo torsion. This means that the real Chern class of $\mathscr{D}$ \eqref{EqCurvDetBd} is a well-defined element of $H^2_{\rm free}(\mathcal{M}/\mathcal{D}^{(1,2)}_\eta, \mathbbm{Z})$. In particular, it has integral periods on surfaces in $\mathcal{M}/\mathcal{D}^{(1,2)}_\eta$. This fact will be important in the following.

\subsection{Quadratic refinements}

\label{SecQuadrRef}

The quantum self-dual field theory on a $4\ell+2$-dimensional manifold with non-trivial cohomology in degree $2\ell+1$ depends on a half-integral element $\eta$ of $H^{2\ell+1}(M,\mathbbm{R})$, which can be interpreted as a theta characteristic. In this section we will introduce another interpretation of $\eta$ that will be of fundamental importance in what follows. Useful references about quadratic refinements include \cite{Brumfiel1973, Taylor1984259, Lee1988}.

Given a finite abelian group $G$ (pictured additively) endowed with a $\mathbbm{Q}/\mathbbm{Z}$-valued non-degenerate symmetric bilinear pairing $L$, a quadratic refinement $q$ of $L$ is a $\mathbbm{Q}/\mathbbm{Z}$-valued function on $G$ satisfying
\be
q(g_1 + g_2) - q(g_1) - q(g_2) = L(g_1,g_2)\;,
\ee
\be
q(ng) = n^2q(g) \;,
\ee
for $g,g_1,g_2 \in G$. Any two quadratic refinements differ by a $\mathbbm{Z}/2\mathbbm{Z}$-valued character of $G$, and all the quadratic refinements can be obtained from a given one in this way. If $G$ admits an isotropic subgroup, i.e. a subgroup $G_0$ on which $L$ vanishes identically, then $q$ is linear on $G_0$ and $2q|_{G_0} = 0$. As a result, by twisting $q$ with an appropriate character, it is always possible to find a quadratic refinement that vanishes on $G_0$. It can be shown (see for instance \cite{Taylor1984259}) that the argument of the Gauss sum
\be
{\rm Gauss}(q) = \sum_{g \in G} \exp \big(2\pi i q(g)\big)
\ee
is a multiple of  $2\pi/8$. The corresponding element $A(q)$ in $\mathbbm{Z}/8\mathbbm{Z}$ is called the generalized Arf invariant of $q$.

One way to obtain the setup described above is to start from a free abelian group $F$ of finite rank endowed with a $\mathbbm{Z}$-valued symmetric pairing $B$. $B$ can be seen as a map from $F$ to its dual $F^\ast$ and we have a short exact sequence
\be
\label{EqSESAbGrp}
0 \rightarrow F \stackrel{B}{\rightarrow} F^\ast \rightarrow G_B \rightarrow 0 \;,
\ee
where $G_B$ is a finite abelian group. As $F \otimes \mathbbm{Q} \simeq F^\ast \otimes \mathbbm{Q}$, $B$ induces a pairing $B^\ast$ on $F^\ast$ which passes to a well-defined pairing $L_B: G_B \times G_B \rightarrow \mathbbm{Q}/\mathbbm{Z}$. An integral Wu class for $B$ is an element $\lambda \in F$ such that $B(w,w) = B(\lambda,w)$ mod $2$ for all $w \in F$. Given an integral Wu class, we can define a quadratic refinement
\be
\label{EqQuadrRefFromWu}
q_\lambda(w+F) := \frac{1}{2}\big(B^\ast(w,w) - B^\ast(w,\lambda)\big) \;\; {\rm mod} \; 1
\ee
of $L_B$. A Wu class is a modulo 2 reduction of an integral Wu class. Wu classes and quadratic refinements are in bijection (Theorem 2.4 \cite{Brumfiel1973}). A theorem of van der Blij \cite{MR0108467} computes the Gauss sum of $q_\lambda$ in terms of $\lambda$: 
\be
\label{EqDeBlijFormula}
{\rm Gauss}(q_\lambda) = |G_B|^{1/2} \exp \frac{2\pi i}{8} \big( \sigma_B  - B(\lambda,\lambda)\big) \;,
\ee
where $\sigma_B$ is the signature of $B$.

We now review how quadratic refinements can be associated to manifolds of dimension $4\ell+2$, $4\ell+3$ and $4\ell+4$.

\paragraph{Dimension $4\ell+2$} Recall that we called $\Lambda$ the free part of the integral cohomology in degree $2\ell+1$ of a $4\ell+2$-dimensional manifold $M$. Set $G = \Lambda/2\Lambda$ and $L = \frac{1}{4\pi}\omega$ mod 1. $\omega$ is antisymmetric, but $L$, being valued in $\frac{1}{2}\mathbbm{Z}/\mathbbm{Z}$, is also symmetric. We will call a quadratic refinement of $L$ a quadratic refinement of the intersection form (QRIF \cite{Belov:2006jd}). Any QRIF can be parameterized as follows
\be
\label{EqRelQRIFChar}
q_\eta(v = v_1 + v^2) = \frac{1}{4\pi} \omega(v_1, v^2) + \frac{1}{2\pi} \omega(\eta, v) \;\; {\rm mod} \; 1\;,
\ee
for $\eta$ a half integral element of $H^{2\ell+1}(M,\mathbbm{R})$. $v_1$ and $v^2$ are the component of $v$ on the two Lagrangian spaces singled out by our choice of Darboux basis in Section \ref{SecIntJacModG}. Hence there is a one to one correspondence between QRIFs and half-integral characteristics, although it depends on a choice of Lagrangian decomposition of $\Lambda$. This is the reason why we restricted our attention to half-integral characteristics in Section \ref{SecIntJacModG}. When $M$ is a framed manifold,\footnote{A framed manifold is a manifold embedded in $\mathbbm{R}^N$ for $N$ large enough, together with a trivialization of its normal bundle.} there is a canonical choice for $q$ and $A(q)$ is a topological invariant of $M$ known as the Kervaire invariant \cite{MR0322888, Snaith}.

\paragraph{Dimension $4\ell+3$} We will also be interested in manifolds $\tilde{M}$ of dimension $4\ell+3$, which will typically be mapping tori of manifolds of dimension $4\ell+2$. In this dimension, the linking pairing defines a symmetric non-degenerate bilinear $\mathbbm{Q}/\mathbbm{Z}$-valued pairing on $H_{\rm tors}^{2\ell+2}(\tilde{M},\mathbbm{Z})$, the torsion subgroup of $H^{2\ell+2}(\tilde{M},\mathbbm{Z})$ \cite{Brumfiel1973}. The definition of the linking pairing is as follows. If $u,v$ are torsion cocycles of degree $2\ell+2$, then there exists $w$ such that $k u = dw$, for $k$ a positive integer and $w$ a cocycle of degree $2\ell+1$. The linking pairing is defined by 
\be
L(u,v) = \frac{1}{k}\langle w \cup v, [\tilde{M}]\rangle \; {\rm mod} \; 1 \;,
\ee 
where $[\tilde{M}]$ is the fundamental class of $\tilde{M}$, $\cup$ the cup product and $\langle \bullet, \bullet\rangle$ is the natural pairing between homology and cohomology. After a choice of quadratic refinement of $L$ is made, its Arf invariant yields a topological invariant, which we will call the Arf invariant of $\tilde{M}$.

\paragraph{Dimension $4\ell+4$} The last case of interest to us is a manifold $W$ of dimension $4\ell+4$ with boundary \cite{Brumfiel1973}. Denote the free quotient of a cohomology group $H^\bullet$ by $H^\bullet_{\rm free}$ and let $F$ be the image of $H_{\rm free}^{2\ell+2}(W,\partial W, \mathbbm{Z})$ into $H_{\rm free}^{2\ell+2}(W, \mathbbm{Z})$. Then the intersection form provides a symmetric pairing $\int_W \bullet \wedge \bullet$ on $F$. Therefore we find ourselves in the situation of \eqref{EqSESAbGrp}, and we obtain a finite abelian group $G_B$ together with a symmetric $\mathbbm{Q}/\mathbbm{Z}$-valued pairing $L_B$. A choice of integral Wu class $\lambda$ provides a quadratic refinement $q_\lambda$ of $L_B$ according to \eqref{EqQuadrRefFromWu}. \eqref{EqDeBlijFormula} then provides a formula for the Arf invariant of $q_\lambda$:
\be
\label{EqDeBlijForm2}
A(q_\lambda) = \sigma_W  - \int_W \lambda^2 \;,
\ee
where we wrote $\sigma_W$ for the signature of the intersection form on $H_{\rm free}^{2\ell+2}(W,\partial W, \mathbbm{Z})$.
To link this case to the one introduced in the previous paragraph, consider a quadratic refinement $q_{\partial W}$ of the linking pairing on $H_{\rm tors}^{2\ell+2}(\partial W,\mathbbm{Z})$. We will call the quadratic refinements $q_\lambda$ and $q_{\partial W}$ compatible if they coincide on the image of
\be
H^{2\ell+2}_{\rm free}(W,\mathbbm{Z}) \rightarrow H_{\rm tors}^{2\ell+2}(\partial W,\mathbbm{Z}) \;.
\ee   
In \cite{Brumfiel1973}, Brumfiel and Morgan showed that for such a pair of compatible quadratic refinements,
\be
\label{EqThBrumMorg}
\frac{1}{8}A(q_\lambda) = - \frac{1}{8} A(q_{\partial W}) + q_{\partial W}(b) \quad {\rm mod} \; 1 \;,
\ee
where $b$ is a certain element of $H_{\rm tors}^{2\ell+2}(\partial W,\mathbbm{Z})$. If $q_{\partial W}$ vanishes on classes in $H_{\rm tors}^{2\ell+2}(\partial W,\mathbbm{Z})$ which come from the restriction of torsion classes in $W$, then the last term in \eqref{EqThBrumMorg} vanishes and the two Arf invariants coincide up to a sign. As the image of $H_{\rm tors}^{2\ell+2}(W,\mathbbm{Z})$ in $H_{\rm tors}^{2\ell+2}(\partial W,\mathbbm{Z})$ is an isotropic subgoup of $G_B$ (\cite{Brumfiel1973}, lemma 3.3), this condition can always be satisfied after a suitable twisting of $q_{\partial W}$. Combining \eqref{EqThBrumMorg} with \eqref{EqDeBlijForm2}, we get the following relation that will be crucial to us:
\be
\label{EqRelTopInv}
A(q_{\partial W}) =  \int_W \lambda^2 - \sigma_W \quad {\rm mod} \; 8 \;.
\ee

\subsection{The Arf invariants of mapping tori}

\label{SecArfMT}

In this section, we compute the Arf invariant of a mapping torus $\hat{M}_c$ associated with the $4\ell+2$-dimensional manifold $M$ and a cycle $c \subset \mathcal{M}/\mathcal{D}^{(2)}$. We essentially follow \cite{Lee1988} and assume for simplicity that $M$ has no torsion cohomology in degree $2\ell+1$. The case where $M$ has torsion is identical, because of a physical condition forcing the quadratic refinement to vanish on the torsion part of $H^{2\ell+1}(M,\mathbbm{Z})$, see Section \ref{SecFormGGA}.

First, we show how the characteristic of the self-dual field theory, or equivalently a QRIF of $M$, determines a quadratic refinement of the linking pairing of $\hat{M}_c$. Recall from Appendix \ref{SecCohomMapTor} that provided $H^{2\ell+1}(M,\mathbbm{Z})$ is torsion-free, the torsion cohomology group of $\hat{M}_c$ in degree $2\ell+2$, called $T^{2\ell+1}(\hat{M}_c)$ in the appendix, is given by 
\be
T^{2\ell+1}(\hat{M}_c) = A/{\rm Im}(1-\phi^\ast) \;,
\ee
where $A$ is the set of elements $v \in H^{2\ell+1}(M,\mathbbm{Z})$ such that $kv = (\mathbbm{1}-\phi^\ast)(w)$ for $w \in H^{2\ell+1}(M,\mathbbm{Z})$ and some non-zero integer $k$. The linking pairing $L_T$ is then the reduction modulo 1 of the following $\mathbbm{Q}$-valued pairing on $A$ \cite{Lee1988}:
\be
\label{EqLinkPairA}
L_A(v_1, v_2) = \frac{1}{2\pi k}\omega(v_1, w_2) \;,
\ee
where $k$ and $w_2$ are chosen such that $kv_2 = (\mathbbm{1}-\phi^\ast)(w_2)$.

We saw in Section \ref{SecWittFormGlobAn} that we can naturally associate a diffeomorphism $\phi_c$ to $c$. $\phi_c$ then determines an element $\gamma_c$ of ${\rm Sp}(2n,\mathbbm{Z})$ through the factorization \eqref{EqProjMCGSp}. Consider a QRIF $q_\eta$ associated with a characteristic $\eta$ as in \eqref{EqRelQRIFChar}. Note that we have $q_\eta(v) = q_\eta(\phi_c^\ast(v)) = q_\eta(\gamma_c \cdot v)$, because we are always restricting ourselves to diffeomorphisms preserving the QRIF. Define $q^\eta_A: A \rightarrow \mathbbm{Q}/\mathbbm{Z}$ as follows
\be
\label{EqLQRefFromChar}
q^\eta_A(v) = \frac{1}{2}L_T(v,v) + q_\eta(v) \quad {\rm mod} \; 1 \;.
\ee
Proposition 2.2.1 of \cite{Lee1988} then shows that $q^\eta_A$ induces on $T^{2\ell+1}(\hat{M}_c)$ a quadratic refinement $q^\eta_T$ of the linking pairing $L_T$.

Given a concrete element $\gamma_c \in \Gamma^{(2)}$, the Arf invariant $A(q^\eta_T)$ is not very difficult to compute. Determine first the set $A$ from the action of $\gamma_c$. Get an expression for the linking pairing $L_T$ from \eqref{EqLinkPairA}, plug the result in \eqref{EqLQRefFromChar} and compute the argument of the corresponding Gauss sum. The results for the set of generators of $\Gamma^{(2)}$ defined in Section \ref{SecTopPicGr} are summarized in table \ref{TableArfMapTor} (see also \cite{Lee1988}).
\begin{table}[htbp]
\renewcommand{\arraystretch}{1.3}
$$\begin{array}{|c||c|c|c|c|c|c|c|}
   \hline
   \gamma_c & \alpha^{(ii)} & \alpha^{(ij)} & \beta^{(ii)} & \beta^{(ij)} & \gamma^{(ii)} & \gamma^{(ij)} & \gamma_{\rm anis} \;  \\
   \hline
   \frac{1}{8} A(q^\eta_T) & 2 \eta_1^i \eta^2_i & 2 \eta_1^i \eta^2_j & \frac{1}{8} - \eta_1^i \eta_1^i & 2 \eta_1^i \eta_1^j & -\frac{1}{8} + \eta^2_i \eta^2_i & 2 \eta^2_i \eta^2_j & 0 \\ \hline
   h \; {\rm mod} \; 2 & 0 & 0 & 1 & 0 & 1 & 0 & 1 \\ 
   \hline
\end{array}$$
\caption{The values of the Arf invariants $A(q^\eta_T)$ of mapping tori associated with elements $\gamma_c$ of the basis of generators of $\Gamma^{(2)}$, as a function of $\eta$. On the third row, we have also listed the dimension modulo 2 of the subspace of $H^{2\ell+1}(M,\mathbbm{R})$ left invariant by $\gamma_c$. The last column provide the corresponding information for an anisotropic transvection $\gamma_{\rm anis} \in \Gamma^{(1,2)}$, whose conjugacy class generates the abelianization of $\Gamma^{(1,2)}$. As $\gamma_{\rm anis}$ leaves only invariant the QRIF corresponding to $\eta = 0$, only the value of $A(q^0_T)$ is displayed in this case.}
\label{TableArfMapTor}
\end{table}
In \cite{Lee1988}, an extremely interesting observation was made, namely that the ratios $A(q^\eta_T)/A(q^{\eta'}_T)$ for a mapping torus $\hat{M}_c$ reproduce the holonomies along $\gamma_c$ of the bundles $\mathscr{F}^{(\eta',\eta)}$, given by the characters \eqref{EqCharCetaCzeroExpl}. It is indeed easy to check that 
\be
\label{EqRelArfTheta}
\exp \frac{2\pi i}{8} \left( A(q^\eta_T) - A(q^{\eta'}_T) \right) =  \chi^{(\eta',0)} \left(\chi^{(\eta,0)}\right)^{-1} \;,
\ee
at least on the system of generators.

According to \eqref{EqRelTopInv}, and provided that the image of $H^{2\ell+2}_{\rm tors}(W,\mathbbm{Z})$ into $H^{2\ell+2}_{\rm tors}(\hat{M}_c,\mathbbm{Z})$ is trivial, $A(q^\eta_T)$ coincides modulo 8 with the topological invariant $\int_W \lambda_\eta^2 - \sigma_W$ of a $4\ell+4$-dimensional manifold $W$ bounded by $\hat{M}_c$, for a choice of integral Wu class $\lambda_\eta$ compatible with $q^\eta_T$. If some of the torsion on $\hat{M}_c$ does arise as the restriction of torsion classes on $W$, then the term $q_{\partial W}(b)$ in \eqref{EqThBrumMorg} does not vanish and we do not have a simple relation between $A(q^\eta_T)$ and $\int_W \lambda_\eta^2 - \sigma_W$.

\section{The global gravitational anomaly}

\label{SecGlobGravAnom}

We now have all the pieces needed to deduce a formula for the global gravitational anomaly of the self-dual field.

\subsection{Taking the fourth root}

\label{SecFourthRoot}

We specialize for now to the case $\eta = 0$. Recall Section \ref{SecHolFormPSD}, where we mentioned that the fourth power of the anomaly bundle $\mathscr{A}^0$ differs from the inverse of the determinant bundle of the signature Dirac operator $\mathscr{D}_s$ by a certain flat bundle $(\mathscr{F}^0)^2$. The holonomy formula for the fourth power of the anomaly bundle is given by
\be
\label{EqHolDs2b}
{\rm hol}_{(\mathscr{A}^0)^4}(c) = (\chi^0(\gamma_c))^2 \exp \pi i (\eta_0 + h) \;,
\ee
where $\chi^0$ is the character generating the torsion group of the Picard group of $\mathcal{T}^{(1,2)}$. According to equation \eqref{EqCharKappa0PB}, $(\chi^0)^2$ is trivially equal to $1$ on elements of $\Gamma^{(2)} \subset \Gamma^{(1,2)}$, but takes the value $-1$ on the anisotropic transvections in $\Gamma^{(1,2)}$ \cite{Monnier2011}. 

As, according to the discussion at the end of Section \ref{SecWittFormGlobAn}, we aim at obtaining a formula involving only topological invariants and the local index density of $\mathscr{D}_s$, this character is clearly an annoyance and we have to get rid of it. It is possible to do so thanks to the following property of the Arf invariant of the mapping torus. Note from Table \ref{TableArfMapTor} that $A(q^0_T) = h$ mod $2$ for all the generators of $\Gamma^{(2)}$, but that for an anisotropic transvection, $A(q^0_T) - h = 1$ mod $2$. The holonomy formula can therefore be rewritten as follows:
\be
\label{EqHolDs2c}
{\rm hol}_{(\mathscr{A}^0)^4}(c) = \exp \pi i \big(\eta_0 - A(q^0_T)\big) \;.
\ee

We now have to take a fourth root of \eqref{EqHolDs2c} to obtain an anomaly formula for $\mathscr{A}^0$. As was already mentioned, this is a non-trivial operation. We will assume the following formula, valid for all $\eta$:
\be
\label{EqHolDs2d}
{\rm hol}_{\mathscr{A}^\eta}(c) = \exp \frac{2\pi i}{8}\big(\eta_0 - A(q^\eta_T)\big) \;.
\ee
Now suppose that $W$ is a manifold bounded by the mapping torus $\hat{M}_c$, such that the image of $H^{2\ell+2}_{\rm tors}(W,\mathbbm{Z})$ into $H^{2\ell+2}_{\rm tors}(\hat{M}_c,\mathbbm{Z})$ is trivial. Using the Atiyah-Patodi-Singer theorem \eqref{EqAPS} and equation \eqref{EqRelTopInv}, we get
\be
\label{EqHolDs2e}
{\rm hol}_{\mathscr{A}^\eta}(c) = \exp \frac{2\pi i}{8} \left( \int_W L_0 - \sigma_W + \sigma_W - \int_W \lambda_\eta^2 \right) = \exp \frac{2\pi i}{8}\int_W \left(L_0 - \lambda_\eta^2 \right) \;.
\ee
In the rest of this section, we will argue that \eqref{EqHolDs2d} and \eqref{EqHolDs2e} do indeed describe the holonomies of the anomaly bundle $\mathscr{A}^\eta$.

\paragraph{The relative anomaly} Let us consider the holonomies of the bundle $\mathscr{A}^{\eta'} \otimes (\mathscr{A}^\eta)^{-1}$, the ``relative anomaly''. We know from \cite{Monnier2011} that topologically, the anomaly bundles $\mathscr{A}^\eta$ are pull-backs from theta bundles over $\mathcal{T}^{(2)}$. As a result, $\mathscr{A}^{\eta'} \otimes (\mathscr{A}^\eta)^{-1} \simeq \mathscr{F}^{(\eta', \eta)}$, which was defined at the end of Section \ref{SecTopPicGr}. The holonomies of $\mathscr{F}^{(\eta', \eta)}$ are described by characters $\chi^{(\eta',0)} \left(\chi^{(\eta,0)}\right)^{-1}$ of $\Gamma^{(2)}$, given explicitly in \eqref{EqCharCetaCzeroExpl}. On the other hand, according to \eqref{EqHolDs2d}, the holonomies of $\mathscr{A}^{\eta'} \otimes (\mathscr{A}^\eta)^{-1}$ are given by 
\be
\label{EqHolDs2f}
{\rm hol}_{\mathscr{A}^{\eta'} \otimes (\mathscr{A}^\eta)^{-1}}(c) = \exp \frac{2\pi i}{8}\big(A(q^\eta_T) - A(q^{\eta'}_T)\big) \;.
\ee
But by \eqref{EqRelArfTheta} this expression is equal to $\chi^{(\eta',0)} \left(\chi^{(\eta,0)}\right)^{-1}$. Therefore, \eqref{EqHolDs2d} reproduces the correct relative anomaly.

\paragraph{The differential character} To show that the holonomy formula defines a well-defined bundle, we have to check that the holonomies define a differential character of degree two on $\mathcal{M}/\mathcal{D}^{(1,2)}_\eta$. (See Section 2 of \cite{Freed:2006yc} for a presentation of the isomorphism between differential characters of degree two and line bundles with connection.) Recall that a differential character of degree two on a manifold $X$ is a homomorphism ${\rm hol}(c)$ from the group of 1-cycles on $X$ into $U(1)$, such that there exists a globally defined 2-form $R$ on $X$ with the following property. For any surface $C$ with boundary $c$, 
\be
\label{EqDefDiffCharact}
{\rm hol}(c) = \exp \int_C R \;.
\ee
Intuitively, this condition is clear: by Stokes theorem, the holonomy along a loop bounding an open surface should be computable from the integral of the curvature over the surface. 

The eta invariant and the Arf invariant are additive under disjoint unions, so ${\rm hol}_{\mathscr{A}^\eta}$ is indeed a homomorphism. From the expression \eqref{EqCurvDetBd} for the local anomaly, we have
\be
R_{\mathscr{A}} = \frac{2\pi i}{8} \left(\int_M L_0\right)^{(2)} \;.
\ee
Comparing with \eqref{EqHolDs2e}, we see that we need to check that $\int_W \lambda_\eta^2$ is a multiple of 8 for each $W$ obtained by restricting $(M \times \mathcal{M})/\mathcal{D}^{(1,2)}_\eta$ to a surface with boundary in $\mathcal{M}/\mathcal{D}^{(1,2)}_\eta$. 

Recall that at the end of Section \ref{SecTopPicGr}, we showed that $\frac{1}{8}\int_M L_0$ has integral periods when integrated over any closed surface in $\mathcal{M}/\mathcal{D}^{(1,2)}_\eta$. Moreover, for $W$ obtained from a closed surface, we have
\be
\int_C \frac{1}{8}\int_M (L_0 - \lambda_\eta^2) = \frac{1}{8}\int_W (L_0 - \lambda_\eta^2) = \frac{1}{8} \left(\sigma_W - \int_W \lambda_\eta^2\right) \in \mathbbm{Z} \;,
\ee
where we used the fact that $L_0$ is the index density of the signature operator, as well as \eqref{EqRelTopInv} in the case where $W$ has no boundary. As a result, we deduce that for $W$ obtained from a closed surface $\int_W \lambda_\eta^2 \in 8\mathbbm{Z}$.

What about open surfaces? As we will see in Section \ref{SecSpinManif}, there are cases in which $\lambda_\eta$ is zero, namely for spin manifolds of dimension $8\ell+2$, for a special choice of QRIF $\hat{\eta}$. Clearly, in these cases, ${\rm hol}_{\mathscr{A}^{\hat{\eta}}}(c)$ defines a differential character of degree 2, and consequently \eqref{EqHolDs2d} and \eqref{EqHolDs2e} describe the holonomies of a well-defined bundle. In general, we do not know how to perform this check. The requirement that $\int_W \lambda_\eta^2 \in 8\mathbbm{Z}$ for all $W$ obtained from open surfaces in $\mathcal{M}/\mathcal{D}^{(1,2)}_\eta$ is a constraint on $\lambda_\eta$. In case no $\lambda_\eta$ would satisfy this constraint, we would interpret it as a sign that our simple choice of fourth roots made in order to obtain \eqref{EqHolDs2d} is not the correct one. This said, it should be noted that the relation to the Hopkins-Singer formalism which we will sketch in Section \ref{SecHopSing} seems to support the choice of fourth roots we made.

In conclusion, except in very special cases, we cannot check that the holonomy formula \eqref{EqHolDs2e} describes the holonomies of well-defined bundles over $\mathcal{M}/\mathcal{D}^{(1,2)}_\eta$. This is the main shortcoming in our derivation. We will nevertheless assume that this is the case in the following.

\paragraph{Modular geometry} Under this assumption, we can show that \eqref{EqHolDs2d} describes the correct fourth root bundle, up to a subtle ambiguity already encountered in \cite{Monnier2011}. Let us temporarily back off and call $\hat{\mathscr{A}}^\eta$ the bundle whose holonomies are given by \eqref{EqHolDs2d}, keeping $\mathscr{A}^\eta$ for the anomaly bundle. We specialize to $\eta = 0$. We know that $(\hat{\mathscr{A}}^0)^4 \simeq (\mathscr{A}^0)^4$. As the dependence on $c$ of $h$, $A(q^0_T)$ and $\chi^0$ factorize through $\gamma_c \in \Gamma^{(1,2)}$, $\hat{\mathscr{A}}^0$ is topologically the pull-back from a bundle on $\mathcal{T}^{(1,2)}$. In fact, $(\mathscr{A}^0)^4$ has exactly four square roots, given by $\mathscr{A}^0 \otimes (\mathscr{F}^0)^i$, $i = 0,1,2,3$. One of them coincides with $\hat{\mathscr{A}}^0$. To eliminate some of these candidates, consider now the bundle $\hat{\mathscr{A}}^\eta$ for an even characteristic $\eta$. $\hat{\mathscr{A}}^\eta$ is the pull-back of a bundle defined over $\mathcal{T}^{(2)}$, but also of a bundle over $\mathcal{T}^{(1,2)}_\eta$. The same is true for $\mathscr{A}^\eta$. If $\mathscr{A}^0 \simeq \hat{\mathscr{A}}^0 \otimes (\mathscr{F}^0)^i$, then 
\be
\mathscr{A}^\eta \simeq \hat{\mathscr{A}}^\eta \otimes (\mathscr{F}^0)^i \;,
\ee
where we used the fact that $\mathscr{A}^\eta \otimes (\mathscr{A}^0)^{-1} \simeq \hat{\mathscr{A}}^\eta \otimes (\hat{\mathscr{A}}^0)^{-1}$, as we checked above. This means that $(\mathscr{F}^0)^i$, as a bundle over $\mathcal{T}^{(2)}$, coincides with the pull-back of a bundle over $\mathcal{T}^{(1,2)}_\eta$. But given the explicit form \eqref{EqCharCeK} of the characters $\chi^\eta$, corresponding to the generator $\mathscr{F}^\eta$ of the torsion of the Picard group of $\mathcal{T}^{(1,2)}_\eta$, we see that this is true only when $i = 0,2$. Indeed, only the trivial bundle and $(\mathscr{F}^\eta)^2$ pull back to $\eta$-independent (actually trivial) bundles on $\mathcal{T}^{(2)}$. Therefore we find that
\be
\label{EqCompAnBHolB}
\mathscr{A}^0 \simeq \hat{\mathscr{A}}^0 \otimes (\mathscr{F}^0)^i \;, \quad i = 0 \; {\rm or} \; 2
\ee
as bundles with connections.

With our current understanding of the self-dual field theory, we cannot hope to resolve the remaining ambiguity in \eqref{EqCompAnBHolB}. Indeed, we lack a complete and unambiguous definition of the self-dual field theory on an arbitrary Riemannian manifold. Techniques such a holomorphic factorization of a non-chiral form \cite{Henningson:1999dm, Dijkgraaf:2002ac} or geometric quantization \cite{Witten:1996hc, Belov:2006jd, Monnier:2010ww}, while they certainly do provide some information about the theory, cannot be considered as satisfactory definitions of the latter, as for instance they do not provide information about the global gravitational anomaly. In \cite{Monnier2011}, our ambition was to perform a first principle derivation, starting from an action principle for the self-dual field. The path integral over this action would have provided a global definition of the partition function, encoding in particular all the information about the global anomaly. We failed in the sense that we have not been able to find an action principle that would yield the correct partition function upon path integration. We offered some heuristic argument about why this failure was in our opinion inevitable in Section 3.4 of \cite{Monnier2011}. However, we have been able to find an action principle for a \emph{pair} of self-dual fields. From the corresponding quantum partition function, we have been able to determine the topological class of the anomaly bundle, up to the same ambiguity as in \eqref{EqCompAnBHolB}. The latter is therefore the unavoidable consequence of the fact that we do not have a good definition of the quantum self-dual field theory on an arbitrary compact oriented Riemannian manifold. At any rate, let us note that the ambiguity is rather mild: if we are willing to consider the anomaly bundle over $\mathcal{M}/\mathcal{D}^{(2)}$, or equivalently consider only diffeomorphisms in $\mathcal{D}^{(2)}$, then the ambiguity disappears. Indeed, $(\mathscr{F}^0)^2$ is a trivial bundle over $\mathcal{M}/\mathcal{D}^{(2)}$, as \eqref{EqCharKappa0PB} shows. In the following, we will always pick what is probably the most natural option: $\mathscr{A}^0 \simeq \hat{\mathscr{A}}^0$. We will see that this choice is also necessary in order for the global gravitational anomaly of type IIB supergravity to vanish.

\subsection{Relation to the Hopkins-Singer bundle}

\label{SecHopSing}

In \cite{Witten:1996hc}, Witten studied the global gauge anomaly of the self-dual field theory coupled to an external gauge field. He showed how the holonomies of the gauge anomaly bundle could be computed by means of a spin Chern-Simons action. In \cite{hopkins-2005-70}, Hopkins and Singer reinterpreted and generalized the ideas of Witten in a purely mathematical context, obtaining a holonomy formula very similar to \eqref{EqHolDs2e}. We will be very sketchy in this section, but it is completely independent of the derivation. Our aim is only to point out the similarities between the two holonomy formulas. We refer the reader to Section 2 of \cite{hopkins-2005-70} for a summary of their work. 

In short, given a map of manifolds $E \rightarrow S$, endowed with a certain extra structure, Hopkins and Singer construct a functor $\kappa$ between certain differential cohomology categories on $E$ and $S$. We are interested in the case when $E$ is a fiber bundle over $S$ with fibers $M$ of dimension $4\ell+2$. Part of the extra structure then describes a family of Riemannian metrics on $M$. The input of the Hopkins-Singer functor is a differential cocycle $\lambda$ lifting the Wu class of the fiber bundle. Admittedly, except in cases where $\lambda$ can be expressed in terms of characteristic classes (see Section \ref{SecSpinManif}), we do not have a prescription to construct $\lambda$ from the quadratic refinement QRIF $\eta$ of $M$. In cases where it is possible to construct such a class $\lambda_\eta$, the functor $\kappa$ produces a differential cohomology class on $S$ which is equivalent to a hermitian line bundle with a unitary connection on $S$. Applying the Hopkins-Singer functor to the fiber bundle $(M \times \mathcal{M})/\mathcal{D}_\eta^{(1,2)}$, we obtain a hermitian line bundle $\mathscr{H\!\!\!S}^\eta$ with connection over $\mathcal{M}/\mathcal{D}_\eta^{(1,2)}$, which we will call the Hopkins-Singer bundle. \footnote{It is not clear whether the Hopkins-Singer theory can be applied directly to bundles over an infinite dimensional base. However, just as it is the case for ordinary anomaly bundles, all the information about the Hopkins-Singer bundle (like its integral Chern class and the holonomies of its connection) can be computed by restricting the fiber bundle to a finite-dimensional base.}

The Hopkins-Singer functor satisfies naturalness conditions that allows one to derive a holonomy formula for the connection on $\mathscr{H\!\!\!S}^\eta$ (see Section 2.7 of \cite{hopkins-2005-70}). Consider again a $4\ell+2$ manifold $M$, and the mapping torus $\hat{M}_c$ associated with a loop $c$ in $\mathcal{M}/\mathcal{D}_\eta^{(1,2)}$. Suppose that there exists a manifold $W$ admitting $\hat{M}_c$ as its boundary. The holonomy of the Hopkins-Singer bundle $\mathscr{H\!\!\!S}^\eta$ is given by
\be
\label{EqHolHS}
{\rm hol}_{\mathscr{H\!\!\!S}^\eta}(c) = \exp \frac{2\pi i}{8} \int_W \left(\lambda_\eta^2 - L \right) \;,
\ee
where $L$ is the Hirzebruch L-genus of $TW$. After taking the adiabatic limit of the Hopkins-Singer formula and comparing with \eqref{EqHolDs2e}, we see that $\mathscr{A}^{\eta} \simeq  (\mathscr{H\!\!\!S}^\eta)^{-1}$ as bundles with connection.

Just like in our construction, the quadratic refinement $q^\eta_{\partial W}$ has to vanish on classes in $H_{\rm tors}^{2\ell+2}(\partial W,\mathbbm{Z})$ which come from the restriction of torsion classes in $W$ (see the proof of proposition 5.44 in \cite{hopkins-2005-70}).

\subsection{A formula for the global gravitational anomaly}

\label{SecFormGGA}

Let us recapitulate the situation in the next two paragraphs, so that the readers who have skipped the derivation can follow us. Recall that the self-dual field on a $4\ell+2$-dimensional manifold $M$ depends on a quadratic refinement of the intersection form (QRIF) on the free quotient $H_{\rm free}^{2\ell+1}(M,\mathbbm{Z})$ of $H^{2\ell+1}(M,\mathbbm{Z})$, which can be represented as a half-integral vector $\eta$ in $H^{2\ell+1}(M,\mathbbm{R})$ modulo $H_{\rm free}^{2\ell+1}(M,\mathbbm{Z})$. When the self-dual field does not couple to an external gauge field, its partition function is the section of a certain line bundle $\mathscr{A}^\eta$ over the space of metrics $\mathcal{M}$ modulo a certain subgroup $\mathcal{D}_\eta^{(1,2)}$ of the group of diffeomorphisms (defined at the end of Section \ref{SecIntJacModG}). The global gravitational anomaly is given by the holonomies of a certain natural connection on $\mathscr{A}^\eta$.

Our aim is to provide a formula computing these holonomies. To this end, consider a loop $c \in \mathcal{M}/\mathcal{D}_\eta^{(1,2)}$. From this loop, we can construct a mapping torus $\hat{M}_c$, which is the restriction to $c$ of the natural fiber bundle $(M \times \mathcal{M})/\mathcal{D}_\eta^{(1,2)}$ over $\mathcal{M}/\mathcal{D}_\eta^{(1,2)}$ to $c$. In order to obtain a practical anomaly formula, we have furthermore to assume that there exists a $4\ell+4$-dimensional manifold $W$ bounded by $\hat{M}_c$ and satisfying a certain technical requirement which we spelled out at the end of Section \ref{SecQuadrRef}. \\

In the preceding sections, we argued that the holonomy of the natural connection on $\mathscr{A}^\eta$ along a loop $c$ in $\mathcal{M}/\mathcal{D}^{(1,2)}_\eta$ is given by the formula \eqref{EqHolDs2e}:
\be
\label{EqHolAB}
{\rm hol}_{\mathscr{A}^\eta}(c) = \exp \frac{2\pi i}{8} \int_W \left(L_0 - \lambda_\eta^2 \right) \;.
\ee
$\lambda_\eta \in H_{\rm free}^{4\ell+2}(W, \mathbbm{Z})$ is an integral Wu-class compatible with the quadratic refinement on the mapping torus induced by $\eta$. We will give more physically relevant information about $\lambda_\eta$ in Sections \ref{SecCombGGA} and \ref{SecSpinManif}, but we should stress that we have an explicit expression for it only in special cases. $L_0$ is the Hirzebruch $L$-polynomial \eqref{EqL-Genus} of the metric on $W$, in the limit where the volume of the fiber of the mapping torus shrinks to zero. $L_0$ is nothing but the local index density used in the computation of the local anomaly of the self-dual field.

\eqref{EqHolAB} is the promised formula for the global anomaly of the self-dual field theory. Provided we can find an expression for $\lambda_\eta$, its structure makes it convenient to check the cancellation of global anomalies, as we already discussed in Section \ref{SecWittFormGlobAn}. In case no suitable manifold $W$ can be found, the holonomy is still described purely in terms of the mapping torus by equation \eqref{EqHolDs2d}, but the latter is probably useless for the purpose of checking anomaly cancellation.

We have to discuss the case when the manifold $M$ has torsion in $H^{2\ell+1}(M,\mathbbm{Z})$. As the pairing on $H^{2\ell+1}(M,\mathbbm{Z})$ vanishes on $H_{\rm tors}^{2\ell+1}(M,\mathbbm{Z})$, the QRIF restricted on $H_{\rm tors}^{2\ell+1}(M,\mathbbm{Z})$ is actually a homomorphism to $\mathbbm{Z}/2\mathbbm{Z}$. If it is non-trivial, then the sum over the zero modes of the self-dual field in the partition function vanishes identically. However, the QRIF can be made trivial on torsion classes by turning on appropriate torsion fluxes on $M$ \cite{Witten:1999vg, Moore:1999gb}. These fluxes should be seen as a necessary ingredient in order to cancel the global gauge anomaly of the self-dual field. As a result, for all practical purposes, the QRIF has to vanish on $H_{\rm tors}^{2\ell+1}(M,\mathbbm{Z})$. Hence it defines a well-defined quadratic refinement on $H_{\rm free}^{2\ell+1}(M,\mathbbm{Z})$ and the formalism developed in Section \ref{SecMathBack} apply. In conclusion, provided one takes care to turn on the appropriate torsion fluxes in order to cancel gauge anomalies, \eqref{EqHolAB} still describes the global gravitational anomaly of the self-dual field when $M$ has torsion in $H^{2\ell+1}(M,\mathbbm{Z})$.

\subsection{The combined gauge-gravitational anomaly}

\label{SecCombGGA}

In this section, we discuss in more detail the physical meaning of the term $\lambda_\eta^2$. We will see that \eqref{EqHolAB} actually describes both the gravitational and gauge global anomalies of the self-dual field theory.

Recall that the self-dual field can be coupled to a background gauge field $\check{C}$. In the holographic description of the self-dual field theory by a spin Chern-Simons theory \cite{Witten:1996hc, Belov:2006jd, Monnier:2010ww}, $\check{C}$ is the restriction of the Chern-Simons $2\ell+1$-form gauge field on the boundary. $\check{C}$ is best described as a differential cocycle of degree $2\ell+2$. \footnote{See Section 2 of \cite{Freed:2006yc} for a pedagogical introduction to differential cocycles, as well as Section 3.1 of \cite{Monnier:2010ww} for details about how to see the spin Chern-Simons field as a differential cocycle.} We fix a flux background on $M$, which is an integral class $a$ in $H^{2\ell+2}(M,\mathbbm{Z})$, to be understood as the class of the field strength of $\check{C}$. The torsion part of $a$ is determined by the cancellation of global gauge anomalies, as explained in the previous section. This choice fixes uniquely a connected component of $\check{H}^{2\ell+2}(M)$. When the gauge transformations of the Chern-Simons theory are taken into account, the space of gauge fields is a torsor $\mathcal{A}$ on $\Omega^{2\ell+1}_{\rm coex}(M) \times \mathcal{J}$, where $\Omega^{2\ell+1}_{\rm coex}(M)$ is the space of co-exact forms on $M$ and $\mathcal{J}$ is the intermediate Jacobian defined in Section \ref{SecIntJacModG}. We write $\mathcal{D}^{(1,2)}_{\eta,a}$ for the subgroup of $\mathcal{D}^{(1,2)}_\eta$ preserving the class $a$. We then form the space $\mathcal{F} := (\mathcal{M} \times \mathcal{A})/\mathcal{D}^{(1,2)}_{\eta,a}$, where $\mathcal{D}^{(1,2)}_{\eta,a}$ acts on $\mathcal{A}$ by pullbacks. $\mathcal{F}$ can be seen as a fiber bundle over $\mathcal{M}/\mathcal{D}^{(1,2)}_{\eta,a}$ with fiber $\mathcal{A}$.

The partition function of the self-dual field is then the section of a line bundle with connection over $\mathcal{F}$. The holonomies of this connection, or equivalently the global anomalies, are associated to cycles $c$ in $\mathcal{F}$. Gravitational anomalies are associated to constant lifts of cycles in $\mathcal{M}/\mathcal{D}^{(1,2)}_{\eta,a}$, while gauge anomalies are computed along cycles in a given fiber $\mathcal{A}$. Mixed anomalies correspond to cycles falling in neither of these two categories. 

We construct as before the mapping torus $\hat{M}_c$ over the cycle $c$ in $\mathcal{F}$. $\hat{M}_c$ comes equipped with a vertical differential cocycle $\check{C}$. We then have to find a manifold $W$ bounded by $\hat{M}_c$, endowed with a compatible integral Wu structure $\hat{\lambda}_\eta$ and on which $\check{C}$ extends. We then choose
\be
\label{EqLambdaGauge}
\lambda_\eta = \hat{\lambda}_\eta - 2G \;,
\ee
where $G$ is the field strength of $\check{C}$ on $W$. 

When the metric is constant along the cycle $c$, or equivalently when $c$ is contained in a single fiber of $\mathcal{F}$, $\hat{M}_c$ reduces to $M \times S^1$, but $\check{C}$ is non-trivial. The anomaly formula \eqref{EqHolAB} gives
\be
\label{EqHolCstM}
{\rm hol}_{\mathscr{A}^\eta}(c) = \exp \left( -\pi i \int_W G \wedge \left(G-\frac{\hat{\lambda}_\eta}{2}\right) \right) \exp \left(\frac{2\pi i}{8} \int_W \left(L_0 - \hat{\lambda}_\eta^2 \right) \right) \;.
\ee
The second factor is the holonomy of the connection on the anomaly bundle along a trivial cycle, so it gives $1$. The first factor is the exponential of the spin Chern-Simons action that is known to compute the global gauge anomaly of the self-dual field \cite{Witten:1996hc}, see also \cite{Belov:2006jd}.

In conclusion, the anomaly formula \eqref{EqHolAB} describes both the global gauge and gravitational anomalies, provided $\lambda_\eta$ is chosen according to \eqref{EqLambdaGauge}. In general there is no universal expression for $\hat{\lambda}_\eta$. However, there exists such an expression when the manifold $M$ is spin.

\subsection{The self-dual field on a spin manifold}

\label{SecSpinManif}

In physical situations, the self-dual field often has to be considered on a spin manifold (the case of the M5-brane is a notable exception). Consider a spin manifold $M$ with a fixed spin structure. Recall that a spin diffeomorphism is a diffeomorphism of $M$ together with a bundle map of the principal bundle characterizing the spin structure (see Section 1 of \cite{Lee1988}). Let us write $\mathcal{D}_s$ for the group of spin diffeomorphisms preserving the given spin structure and the class $a$. As in the previous section, we can form the quotient
\be
\label{EqDefFs}
\mathcal{F}_s := (\mathcal{M} \times \mathcal{A})/\mathcal{D}_s \;.
\ee
The construction of the mapping torus then goes as before.

The main difference with the non-spin case is that the spin structure determines canonically a QRIF $\hat{\eta}$, although an explicit form for the latter is in general difficult to obtain \cite{Witten:1999vg}. What can be obtained explicitly is the corresponding class $\hat{\lambda}_{\hat{\eta}}$ (see Appendix E of \cite{hopkins-2005-70}). For $M$ of dimension $8\ell+2$, $\hat{\lambda}_{\hat{\eta}}$ can be taken to vanish, while in the dimension 6 case, $\hat{\lambda}_{\hat{\eta}} = - \frac{1}{2}p_1$, where $p_1$ is the first Pontryagin class. In the case of the M5-brane, the matter is a bit more complicated  \cite{Witten:1996hc, Witten:1999vg}. 

In the case of the spin dimension 6 clase, let us note that it is not quite clear that $\hat{\lambda}_{\hat{\eta}} = - \frac{1}{2}p_1$ is compatible with the known local anomaly of the self-dual field. Indeed, recall that in Section \ref{SecFourthRoot}, we saw that $\frac{1}{8} \int_W \hat{\lambda}_{\hat{\eta}}$ has to be an integer on any $W$ which is obtained by restriction the fibration $(M \times \mathcal{M})/\mathcal{D}_s$ to a surface with boundary on $\mathcal{M}/\mathcal{D}_s$. For the Ansatz $\hat{\lambda}_{\hat{\eta}} = - \frac{1}{2}p_1$, we can check that this is true on manifolds $W$ obtained from closed surfaces, but we do not know if this holds when the surface has a boundary. These checks should be performed before trying to apply \eqref{EqHolAB} to the six-dimensional self-dual field.

\section{Discussion}

\label{SecDissc}

We explain here why our derivation is not rigorous and discuss the limitations of the anomaly formula. We offer a conjecture summarizing the mathematical insight we gained. We also show that type IIB supergravity as it was understood in the 80's is free of global gravitational anomalies. Finally, we offer an outlook about other physical applications.

\subsection{About the derivation}

\label{SecDeriv}

We list here the points where our fails to be rigorous. We should emphasize that we have done our best to list all the possible shortcomings. We do not believe any of them is very serious except for the last one.
\begin{enumerate}
  \item The Dirac operator on the mapping torus that should enter the Bismut-Freed formula is not quite the Atiyah-Patodi-Singer operator (see the footnote near equation \eqref{Eq4l3DO}). Consider a generic Dirac operator $D$ on $M$, twisted by a certain bundle $\mathscr{T}$. The Bismut-Freed prescription to construct the Dirac operator $\hat{D}$ on the mapping torus $\hat{M}_c$ is to take the untwisted Dirac operator on $\hat{M}_c$ and twist it with $\mathscr{T}$, seen as a bundle over $\hat{M}_c$. In the case of the signature operator, the twist bundle is the spin bundle $\mathscr{S}$ of $M$, seen as a bundle over $\hat{M}_c$. On the other hand, the twist bundle associated to the Atiyah-Patodi-Singer operator $\hat{D}_s$ is the spin bundle $\hat{\mathscr{S}}$ of $\hat{M}_c$. While these
bundles coincide topologically, the connections they carry differ. The connection on $\hat{\mathscr{S}}$ is induced by the Levi-Civita connection on $\hat{M}_c$, while the connection on $\mathscr{S}$ only involves the Levi-Civita connection of the vertical tangent bundle. Some components of the Levi-Civita connection on $\hat{M}_c$, arising because the vertical metric is not constant along the loop, do not appear in the connection on $\mathscr{S}$. As a result, the Dirac operators on $\hat{M}_c$ constructed by twisting the ordinary Dirac operator by either $\hat{\mathscr{S}}$ or $\mathscr{S}$ differ, although they tend to the same limit when $\epsilon \rightarrow 0$. It is believed that the holonomies computed by using either operator in the Bismut-Freed formula agree in this limit, but this has apparently not been proven.
	\item The information about the Picard groups of $\Gamma^{(2)}$ and of $\Gamma^{(2,1)}$ that we presented in Section \ref{SecTopPicGr}, which is central to our argument in the present paper and in \cite{Monnier2011}, is valid only for $n \geq 3$, where $2n$ is the dimension of $H^{2\ell+1}(M,\mathbbm{R})$. Strictly speaking we cannot guarantee that the anomaly formula takes the same form when $n \leq 2$. This puts in principle a restriction on the type of manifolds on which we can check anomaly cancellation. On the other hand, one can argue that if the anomaly formula was indeed taking a different form for $n \leq 2$, this would be likely to impair the nice anomaly cancellation in type IIB supergravity that we derive in Section \ref{SecGlobAnCan}. Note that our formula correctly describes the case $n = 0$, where the anomaly bundle is a square root of $\mathscr{D}^{-1}$.
	\item It is clear that the group $\Gamma^{(1,2)}_\eta$ leaving fixed $\eta$ for an even characteristic is isomorphic to $\Gamma^{(1,2)}$. It is in fact a conjugate subgroup in ${\rm Sp}(2n,\mathbbm{Z})$. However, for odd characteristics, $\Gamma^{(1,2)}_\eta$ might be a different group. In consequence, the parts of our argument where we use the character group of $\Gamma^{(1,2)}_\eta$ are valid only for even characteristic. We are not aware of a computation of the abelianization of the theta group for odd characteristics. 
	\item We identify the character $(\chi^0)^2$ with $\exp \pi i \big(h - A(q^0_T)\big)$ in order to obtain \eqref{EqHolDs2c}. We check that this identity holds for the elements of $\Gamma^{(2,1)}$ appearing in Table \ref{TableArfMapTor}, but we do not know of a proof of this fact valid for the whole group.  In \cite{Lee1988}, it seems that similar statements are checked only on the generators of the relevant group. This would be fine if we knew that both sides of the equality were characters, but this is not the case for the side involving the Arf invariant. Of course, the statement can be tested for any given group element by a direct computation and we always found equality.
		\item Except in very special cases, we have not been able to prove that the choice \eqref{EqHolDs2d} of fourth root of holonomies defines a line bundle over $\mathcal{M}/\mathcal{D}^{(1,2)}_\eta$. This problem is explained in detail in the paragraph entitled ``The differential character'' of Section \ref{SecFourthRoot}. This is by far the most important shortcoming in our opinion.
\end{enumerate}
In addition, the following facts should be kept in mind when trying to apply the anomaly formula to a physical problem.
\begin{enumerate}
	\item We faced a mild ambiguity when we tried to identify the anomaly bundle. We already discussed this issue in detail at the end of Section \eqref{SecFourthRoot}. In summary, we know that our anomaly formula describes the holonomies of the anomaly bundle twisted by a character of order two of $\Gamma^{(1,2)}$, but we do not know if this character is trivial or not. (Recall that $\Gamma^{(1,2)}$ admits only a single non-trivial character of order two.) This ambiguity is due to the lack of a complete definition of the self-dual field theory on an arbitrary Riemannian manifold, and we cannot hope to solve it with our current knowledge. Nevertheless, in our opinion the fact that we obtain perfect anomaly cancellation for type IIB supergravity in Section \ref{SecGlobAnCan} by assuming a trivial character settles the matter for all practical purposes.
		\item As was already mentioned above, $\Gamma^{(1,2)}_\eta$ for an odd characteristic $\eta$ could admit more than a single non-trivial character of order two and the ambiguity might be larger than only two-fold. Note that it was conjectured in \cite{Henningson:2010rc} that the characteristics determined by a spin structure are always even.
	\item In order to obtain useful holonomy formulas for the anomaly bundle, we assume that the mapping torus $\hat{M}_c$ under consideration is bounding a $4\ell+4$-dimensional manifold $W$ such that the quadratic refinements on $\hat{M}_c$ vanishes on classes coming from the restriction of torsion classes on the bounded manifold $W$. The relevant cobordism groups and spectra have been described in Section 5.1 of \cite{hopkins-2005-70}, but to our knowledge they have not been computed explicitly. As a result, it is possible that some of the mapping tori with quadratic refinement that we consider do not bound a suitable manifold $W$. In principle, this implies that one cannot compute the global anomalies associated with certain cycles by means of \eqref{EqHolAB}. The holonomy is still given by \eqref{EqHolDs2d}, but it is probably impossible to check anomaly cancellation with this formula.
	\item The formula can be used practically only when there exists a universal expression for $\lambda_\eta$ in terms of characteristic classes, which happens for spin manifolds. We lack a general construction for $\lambda_\eta$.
	\item Compatibility with the local anomaly \eqref{EqCurvDetBd} puts constraints on $\lambda_\eta$ as explained in Sections \ref{SecFourthRoot} and \ref{SecSpinManif}. It is not obvious to us that the Ansatz for six dimensional spin manifolds described in Section \ref{SecSpinManif} satisfies them.
\end{enumerate}

\subsection{A mathematical conjecture and a question}

\label{SecMathConj}

Although our derivation is not rigorous, we believe we identified an important relation between the Hopkins-Singer functor, Siegel theta functions and the index theory of Dirac operator. In this section, we would like to summarize them briefly. 

We use the notations of Section \ref{SecMathBack}. Consider a fiber bundle $E \rightarrow S$ with fiber $M$ of dimension $4\ell+2$, endowed with an $\check{H}$-orientation in the sense of Hopkins-Singer (see Section 2.4 of \cite{hopkins-2005-70}). Let $\eta$ be a quadratic refinement of the intersection form mod 2 on $H^{2\ell+1}(M,\mathbbm{Z})$ that vanishes on $H_{\rm tors}^{2\ell+1}(M,\mathbbm{Z})$. Suppose moreover that the monodromy action of the fiber bundle preserves $\eta$.

\paragraph{Question} From the data above, is it possible to use the Hopkins-Singer functor to construct a line bundle $\mathscr{H\!\!\!S}^\eta$ over $S$, whose holonomies are computed by \eqref{EqHolDs2d} ? \\

To answer this question by the affirmative, one has to make precise the discussion at the end of Section \ref{SecHopSing}, and in particular to construct the compatible integral Wu structure $\lambda_\eta$ on $E$ associated to the quadratic refinement $\eta$. $\lambda_\eta$ will have to satisfy the constraints exposed in Section \ref{SecFourthRoot}. When $M$ is spin of dimension $8\ell+2$, we know that $\lambda_\eta = 0$ is a solution for a particular quadratic refinement $\hat{\eta}$ (see Section \ref{SecSpinManif}).

There is another way to construct a bundle over $S$ from the data available. The Hodge star operators associated to the family of Riemannian metrics on the fibers of $E$, when restricted to the space of harmonic forms of degree $2\ell+1$, define a period map into the space $\mathcal{T}^{(1,2)}_\eta$ (see Section \ref{SecIntJacModG} for notations). Over $\mathcal{T}^{(1,2)}_\eta$, there exists a theta bundle $\mathscr{C}^\eta$, whose factor of automorphy is given by the multiplier of the theta constant with characteristic $\eta$. We can pull-back this bundle to a bundle $\tilde{\mathscr{C}^\eta}$ on $S$.

\paragraph{Conjecture} When the construction of $\mathscr{H\!\!\!S}^\eta$ described in the question above is possible, as topological bundles, $\mathscr{H\!\!\!S}^\eta \simeq (\tilde{\mathscr{C}}^\eta)^{-1}$. \\

It is clear that the local form of the curvature of the connection on $(\mathscr{H\!\!\!S}^\eta)^2$ coincides locally with the Bismut-Freed connection on the determinant bundle of the Dirac operator coupled to chiral spinors (see Sections \ref{SecSetup} and \ref{SecLocAnom}). The latter remark and the conjecture characterize the Hopkins-Singer bundle with its connection in terms of Siegel theta bundles and index theory.

These statements should also have generalizations when the fiber bundle is endowed with a vertical differential cocycle $x$ of degree $2\ell+2$. Then the $\mathbbm{R}$-valued cocycle of degree $2\ell+1$ which is part of $x$ defines a map from $S$ into a shifted intermediate Jacobian $\mathcal{J}$. By considering the Hodge star operator as above, we get a map from $S$ into the universal abelian variety over $\mathcal{T}^{(1,2)}_\eta$. The theta bundle is defined over this abelian variety, so we can pull it back to $S$ and obtain again a bundle $\tilde{\mathscr{C}}^\eta$. We expect the conjecture to be valid as well in this extended context.

\subsection{Global anomaly cancellation in type IIB supergravity}

\label{SecGlobAnCan}

As a simple application of our results, we check that type IIB supergravity is free of gravitational anomalies, along the lines of the original paper of Witten \cite{Witten:1985xe}. We should stress that in view of the knowledge gained about type IIB supergravity and string theory in the last two decades, our check is naive. Indeed, it ignores the fact that the gauge fields of type IIB supergravity are differential K-theory classes\footnote{Thanks to Greg Moore for reminding us about this.} \cite{Moore:1999gb}, as well as the subtleties associated to non-trivial background fluxes. We hope to offer a more complete treatment of the cancellation in a future paper. At any rate, it provides a simple example of how the anomaly formula can be applied. It also suggests that the correct resolution of the ambiguity encountered in the determination of the anomaly bundle consists in choosing the trivial character of $\Gamma^{(2,1)}$ (see the end of Section \ref{SecFourthRoot}).


Consider a 10 dimensional spin manifold $M$ with spin structure $s$. Let $\tilde{\eta}$ be the characteristic associated to this data by the procedure explained in \cite{Witten:1996hc}. Let $\mathcal{D}_s$ be the subgroup of diffeomorphism preserving the characteristic $\tilde{\eta}$ as well as the spin structure $s$. Let us pick a cycle $c$ in $\mathcal{M}/\mathcal{D}_s$ and construct the corresponding mapping torus $\hat{M}_c$.

We first compute the global anomalies along $c$ of the chiral Dirac field and of the Rarita-Schwinger field on a 10 dimensional spin manifold $M$. We write $\mathscr{D}_D$ for the determinant bundle of the Dirac operator $D_D$ on $M$ and $\mathscr{D}_R$ for the Rarita-Schwinger operator $D_R$ (the Dirac operator twisted by the tangent bundle of $M$). We equip them with their natural Bismut-Freed connections. As in Section \ref{SecWittFormGlobAn}, we resize the metric on the circle in $\hat{M}_c$ by a factor $\frac{1}{\epsilon^2}$. We denote by $\eta_D(\epsilon)$ the eta invariant of the Dirac operator on $\hat{M}_c$ and $h_D(\epsilon)$ the dimension of its kernel. Similarly, we write $\eta_S(\epsilon)$ for the eta invariant of the Rarita-Schwinger operator on $\hat{M}_c$ and $h_R(\epsilon)$ for the dimension of its kernel. It is useful to define the quantities 
\be
\xi_D(\epsilon) := \frac{1}{2}\big(\eta_D(\epsilon) + h_D(\epsilon)\big) \quad {\rm and} \quad \xi_R(\epsilon) := \frac{1}{2}\big(\eta_R(\epsilon) + h_R(\epsilon)\big) \;.
\ee
Using the Bismut-Freed formula, we obtain the following holonomy formulas
\begin{align}
{\rm hol}_{\mathscr{D}_D}(c) = &\; \lim_{\epsilon \rightarrow 0} \; (-1)^{{\rm index}(D_D)} \exp -2\pi i \xi_D(\epsilon) \;,\\
{\rm hol}_{\mathscr{D}_R}(c) = &\; \lim_{\epsilon \rightarrow 0} \; (-1)^{{\rm index}(D_R)} \exp -2\pi i (\xi_R(\epsilon) - \xi_D(\epsilon)) \;.
\end{align}
The second formula comes from the fact that the tangent space of $\hat{M}_c$ decomposes into the direct sum of the tangent space of $M$ and a one-dimensional vector space. We now assume that $\hat{M}_c$ bounds a 12-dimensional manifold $W$ whose metric extends the metric on $\hat{M}_c$ and looks like a direct product near $\hat{M}_c$. We use the Atiyah-Patodi-Singer theorem to reexpress the eta invariants \cite{Witten:1985xe}:
\begin{align}
\lim_{\epsilon \rightarrow 0} \xi_D(\epsilon) = &\; \int_W \hat{A}(R_0) - {\rm index} D^{(W)}_D \;, \\
\lim_{\epsilon \rightarrow 0} \xi_R(\epsilon) = &\; \int_W \big(K(R_0) - \hat{A}(R_0)\big) - {\rm index} D^{(W)}_R + {\rm index} D^{(W)}_D \;,
\end{align} 
where $R_0$ is the curvature of $W$ in the limit $\epsilon \rightarrow 0$. $D^{(W)}_D$ and $D^{(W)}_R$ are respectively the Dirac and Rarita-Schwinger operators on $W$ and $\hat{A}(R_0)$ and $K(R_0)$ are their index densities.

The holonomy of the bundle associated with the self-dual field is given by \eqref{EqHolAB}. Moreover, we know that we can choose $\lambda_{\tilde{\eta}} = 0$ in the 10 dimensional spin case we are considering \cite{Witten:1996hc}. We get therefore:
\be
\label{EqHolSDIIB}
{\rm hol}_{\mathscr{A}^\eta}(c) = \exp 2\pi i \int_W \frac{1}{8}L_0 \;.
\ee

Let us consider the type IIB supergravity multiplet. The bosonic fields are the metric, two scalars, two 2-forms gauge fields and a self-dual 4-form gauge field. The fermionic field content is given by two Majorana-Weyl (spin $\frac{1}{2}$) fermions of negative chirality and two Majorana Rarita-Schwinger (spin $\frac{3}{2}$) fermions of positive chirality. The anomalous fields are the self-dual field and the fermions. The anomaly of the pair of spin $\frac{1}{2}$ fermions is described by means of the index theory of the Dirac operator:
\be
\label{EqHolDir}
{\rm hol}_{\mathscr{D}^{-1}_D}(c) = \exp 2\pi i \int_W \hat{A}(R_0)\;,
\ee
while the Rarita-Schwinger operator is relevant to the pair of spin $\frac{3}{2}$ fermions:
\be
\label{EqHolRS}
{\rm hol}_{\mathscr{D}_R}(c) = \exp -2\pi i \int_W \big(K(R_0) - \hat{A}(R_0)\big)\;,
\ee
The two pairs of Majorana fermions can be seen as two Weyl fermions, which is why we can use the holonomy formulas for $\mathscr{D}_D$ and $\mathscr{D}_R$. We computed the holonomy of $\mathscr{D}^{-1}_D$ rather than $\mathscr{D}_D$ because the spin $\frac{1}{2}$ fermions have negative chirality. Also, it is clear that the factors proportional to the various indices of Dirac operators do not contribute. Multiplying \eqref{EqHolSDIIB}, \eqref{EqHolDir} and \eqref{EqHolRS}, we obtain the sum of the index densities associated to the anomalous fields. But the sum is zero because the local anomaly vanishes \cite{AlvarezGaume:1983ig}. In consequence, the global anomaly vanishes as well, for any choice of 10-dimensional spin manifold $M$. 

The reader might be surprised that we do not recover Witten's formula \eqref{EqHolNZModes4}. Using the latter, we would have deduced that the global anomaly of type IIB supergravity is given by $\exp -2\pi i \sigma_W/8$ \cite{Witten:1985xe}, obtaining an apparently non-vanishing global anomaly. $\sigma_W$ mod $16$ is known as the Rohlin invariant of $\hat{M}_c$. The puzzle is solved by the fact that when $H^{2\ell+1}(M,\mathbbm{Z}) = 0$, the Rohlin invariant is always a multiple of $8$. This can be understood from \eqref{EqRelTopInv}, which was derived from the work of Brumfiel and Morgan \cite{Brumfiel1973} (see also \cite{Lee1988}). Indeed, in this case the torsion group $T^{2\ell+1}$ of the mapping torus constructed from $M$ is necessarily trivial (see appendix \ref{SecCohomMapTor}), so the corresponding Arf invariant vanishes. \eqref{EqRelTopInv} then states that $\sigma_W = 0$ mod $8$. In conclusion, our formula does coincide with Witten's formula in case $H^{2\ell+1}(M,\mathbbm{Z}) = 0$, and the latter does not predict any anomaly for type IIB supergravity in the cases where it can be applied.

\subsection{Other physical applications}

\label{SecOthPhysApp}

Along the same lines, it would be interesting to check global anomaly cancellation in six-dimensional supergravities containing self-dual fields (see for instance \cite{Avramis:2006nb}). Note that the matter will not be quite as simple as in 10 dimensions, as the class $\hat{\lambda}_\eta$ cannot be taken to vanish anymore, and the intersection product on the bounded 8 dimensional manifold will not necessarily be even. See also the puzzle described at the end of Section \ref{SecCombGGA}.

Anomaly cancellation of six dimensional supergravities is especially important in the context of the exploration of the landscape of six-dimensional compactifications of string and M-theory (see \cite{Taylor:2011wt} for a review). The two main aims of this program is to figure out which six-dimensional effective field theory can be consistently coupled to gravity, and which ones can be realized as a compactification of a string theory or of M-theory. In particular, one would like to know if these two sets coincide, which would reveal a striking universality of the string/M-theory construction. One requirement for a theory to couple consistently to gravity is that it should be free of gravitational anomalies, and checking for anomaly cancellation allows one to eliminate a large number of candidates. One can then try to realize the surviving candidates explicitly as compactifications. Typically, the relevant field theories involve self-dual fields in six dimensions, which made it impossible to use global gravitational anomaly cancellation as a constraint. We can now hope to use this criterion and make further steps toward closing the gap between apparently consistent effective field theories and theories which can actually be realized as compactifications.

We already mentioned in the previous section the problem of checking global anomalies for type IIB supergravity backgrounds, but self-dual fields also play a role in M-theory backgrounds containing M5-branes, as well as IIA and heterotic $E_8 \times E_8$ backgrounds containing NS5-branes. A global anomaly formula for the five-brane worldvolume theory would be desirable. In addition to solving the puzzle associated with the six dimensional self-dual field, deriving such an anomaly formula will require a careful consideration of the couplings to the bulk fields.

Another field which should benefit from these results is the computation of instanton correction to low energy supergravity by five-brane instantons (see for instance \cite{Becker:1995kb, Witten:1996bn, Dijkgraaf:2002ac, Tsimpis:2007sx, Donagi:2010pd, Alexandrov:2010ca}), either in M-theory or in type IIA string theory. Indeed, the presence of anomalies restrains the type of instantons that can contribute. While there exist criteria as to whether an instanton will contribute or cannot contribute to the low energy effective action \cite{Witten:1996bn}, a necessary and sufficient criterion remains to be found. The consideration of global anomalies should allow progresses toward its elaboration.

We hope to return to these questions in the future.

\subsection*{Acknowledgments}

I would like to thank Dan Freed and Greg Moore for correspondence and discussions. Part of this work was done while I was visiting the New High Energy Theory Center at the Physics department of Rutgers University, which I would like to thank for hospitality and generous financial support. This work was supported by a Marie Curie intra-European fellowship, grant agreement number 254456.

\appendix

\section{The cohomology of mapping tori}

\label{SecCohomMapTor}

In this appendix, we compute the cohomology of a mapping torus $\hat{M}_\phi$ associated with a diffeomorphism $\phi$ of a manifold $M$ of dimension $4\ell+2$. We also compute more explicitly the torsion cohomology group in degree $2\ell+1$. Finally, we provide a formula for the linking pairing on the torsion group. Part of this material can be found in Section 2 of \cite{Lee1988}.

Recall that $\hat{M}_\phi$ is defined as $M \times [0,1]$, subject to the identification $(x,0) \sim (\phi(x),1)$. As a result, $\hat{M}_\phi$ is a fiber bundle over a circle $S^1$, with fiber $M$. The cohomology can be computed from the following long exact sequence (see for instance Example 2.48 in \cite{Hatcher_2002} or Section 4.1 in \cite{2010arXiv1003.5084C}):
\begin{align}
\label{EqSeqCohMT}
0 \rightarrow & \, H^1(\hat{M}_\phi,\mathbbm{Z}) \stackrel{i^\ast}{\rightarrow} H^1(M,\mathbbm{Z}) \stackrel{1-\phi^\ast}{\rightarrow} H^1(M,\mathbbm{Z}) \stackrel{\theta \cup}{\rightarrow} \\
& H^2(\hat{M}_\phi,\mathbbm{Z}) \stackrel{i^\ast}{\rightarrow} H^2(M,\mathbbm{Z}) \stackrel{1-\phi^\ast}{\rightarrow} H^2(M,\mathbbm{Z}) \stackrel{\theta \cup}{\rightarrow} H^3(\hat{M}_\phi,\mathbbm{Z}) ... \notag
\end{align}
where $i$ is the inclusion of the fiber of $\hat{M}_\phi$, $\phi^\ast$ is the action of the diffeomorphism by pull-back and $\theta$ is the pull-back to $\hat{M}_\phi$ of the generator of $H^1(S^1,\mathbbm{Z})$.

On the first line, as the kernel of $1-\phi^\ast$ is $H_\phi^1(M,\mathbbm{Z})$, the invariant subspace of $H^1(M,\mathbbm{Z})$ under $\phi^\ast$, we deduce that $H^1(\hat{M}_\phi,\mathbbm{Z}) \simeq H_\phi^1(M,\mathbbm{Z})$. If we consider the sequence \eqref{EqSeqCohMT} over $\mathbbm{R}$, the image of $1-\phi^\ast$ in $H^1(M,\mathbbm{R})$ is the complement of $H_\phi^1(M,\mathbbm{R})$. Over $\mathbbm{Z}$, the matter is a bit more subtle as ${\rm Im}(1-\phi^\ast)$ can be a sublattice of $A := H^1(M,\mathbbm{Z})/H_\phi^1(M,\mathbbm{Z})$. As a result, the image of $\theta \cup$ in $H^2(\hat{M}_\phi,\mathbbm{Z})$ is $H_\phi^1(M,\mathbbm{Z}) \oplus T^1(\hat{M}_\phi)$, where $T^1(\hat{M}_\phi)$ is the torsion group $A/{\rm Im}(1-\phi^\ast)$. Repeating the first part of the argument above, we deduce that $H^2(\hat{M}_\phi,\mathbbm{Z}) \simeq H_\phi^1(M,\mathbbm{Z}) \oplus H_\phi^2(M,\mathbbm{Z}) \oplus T^1(\hat{M}_\phi)$. In general, we have
\be
H^p(\hat{M}_\phi,\mathbbm{Z}) \simeq H_\phi^{p-1}(M,\mathbbm{Z}) \oplus H_\phi^p(M,\mathbbm{Z}) \oplus T^{p-1}(\hat{M}_\phi) \;,
\ee
where $T^p(\hat{M}_\phi)$ is the quotient of $H^{p-1}(M,\mathbbm{Z})/H_\phi^{p-1}(M,\mathbbm{Z})$ by the image of $1-\phi^\ast$. Of course, the real cohomology is given by 
\be
H^p(\hat{M}_\phi,\mathbbm{R}) \simeq H_\phi^{p-1}(M,\mathbbm{R}) \oplus H_\phi^p(M,\mathbbm{R}) \;.
\ee
We can now compute the parity of the dimension of the kernel of the Atiyah-Patodi-Singer Dirac operator. Recall that the latter acts on the even forms of $\hat{M}_\phi$ and that its zero modes are simply the harmonic forms. $\phi$ leaves invariant the top class in $H^{4\ell+2}(M,\mathbbm{R})$. It follows from the non-degeneracy of the pairing on $H^\bullet(M,\mathbbm{R})$ that to each invariant class in $H_\phi^p(M,\mathbbm{R})$, there exists an invariant class in $H_\phi^{4\ell+2-p}(M,\mathbbm{R})$. As $H^{\rm even}(\hat{M}_\phi,\mathbbm{R}) \simeq H_\phi^\bullet(M,\mathbbm{R})$, the parity of its dimension is equal to the parity of the dimension of $H_\phi^{2\ell+1}(M,\mathbbm{R})$. This is the result we used in Section \ref{SecWittFormGlobAn} and in Table \ref{TableArfMapTor}.

We now give a formula for the linking pairing on the torsion group $T^{2\ell+1}(\hat{M}_\phi)$. Let $A$ be the set of elements $v \in H^{2\ell+1}(M,\mathbbm{Z})$ such that $kv = (\mathbbm{1}-\phi^\ast)(w)$ for $w \in H^{2\ell+1}(M,\mathbbm{Z})$ and some non-zero integer $k$. Then 
\be
T^{2\ell+1}(\hat{M}_\phi) = A/{\rm Im}(1-\phi^\ast) \;.
\ee
Consider the following $\mathbbm{Q}$-valued pairing on $A$:
\be
L_A(v_1, v_2) = \frac{1}{k}I(v_1, w_2) \;,
\ee
where $I$ is the intersection product on $H^{2\ell+1}(M,\mathbbm{Z})$, and $k$ and $w_2$ are chosen such that $kv_2 = (\mathbbm{1}-\phi^\ast)(w_2)$. Then it is shown in Proposition 2.1.1 of \cite{Lee1988} that $L_A$ induces a well defined $\mathbbm{Q}/\mathbbm{Z}$-valued pairing $L_T$ on $T^{2\ell+1}(\hat{M}_\phi)$ and that it coincides with the linking pairing.

%

{
\small

\begin{thebibliography}{10}

\bibitem{Monnier2011}
S.~Monnier, ``The anomaly bundle of the self-dual field theory'',
  \href{http://arXiv.org/abs/1109.2904}{{\tt 1109.2904}}.

\bibitem{Witten:1985xe}
E.~Witten, ``Global gravitational anomalies'', {\em Commun. Math. Phys.} {\bf
  100} (1985)
197.

\bibitem{Moore:1999gb}
G.~W. Moore and E.~Witten, ``Self-duality, {R}amond-{R}amond fields, and
  {K}-theory'', {\em JHEP} {\bf 05} (2000) 032,
\href{http://arXiv.org/abs/hep-th/9912279}{{\tt hep-th/9912279}}.

\bibitem{Diaconescu:2000wy}
D.-E. Diaconescu, G.~W. Moore, and E.~Witten, ``E(8) gauge theory, and a
  derivation of {K}-theory from {M}-theory'', {\em Adv. Theor. Math. Phys.}
  {\bf 6} (2003) 1031--1134,
\href{http://arXiv.org/abs/hep-th/0005090}{{\tt hep-th/0005090}}.

\bibitem{Evslin:2006cj}
J.~Evslin, ``What does(n't) {K}-theory classify?'',
\href{http://arXiv.org/abs/hep-th/0610328}{{\tt hep-th/0610328}}.

\bibitem{Henningson:1997da}
M.~Henningson, ``{Global anomalies in M-theory}'', {\em Nucl. Phys.} {\bf B515}
  (1998) 233--245,
\href{http://arXiv.org/abs/hep-th/9710126}{{\tt hep-th/9710126}}.

\bibitem{Becker:1999kh}
K.~Becker and M.~Becker, ``{Fivebrane gravitational anomalies}'', {\em Nucl.
  Phys.} {\bf B577} (2000) 156--170,
\href{http://arXiv.org/abs/hep-th/9911138}{{\tt hep-th/9911138}}.

\bibitem{Witten:1996hc}
E.~Witten, ``{Five-brane effective action in M-theory}'', {\em J. Geom. Phys.}
  {\bf 22} (1997) 103--133,
\href{http://arXiv.org/abs/hep-th/9610234}{{\tt hep-th/9610234}}.

\bibitem{Becker:1995kb}
K.~Becker, M.~Becker, and A.~Strominger, ``{Fivebranes, membranes and
  nonperturbative string theory}'', {\em Nucl. Phys.} {\bf B456} (1995)
  130--152,
\href{http://arXiv.org/abs/hep-th/9507158}{{\tt hep-th/9507158}}.

\bibitem{Witten:1996bn}
E.~Witten, ``{Non-Perturbative Superpotentials In String Theory}'', {\em Nucl.
  Phys.} {\bf B474} (1996) 343--360,
\href{http://arXiv.org/abs/hep-th/9604030}{{\tt hep-th/9604030}}.

\bibitem{Dijkgraaf:2002ac}
R.~Dijkgraaf, E.~P. Verlinde, and M.~Vonk, ``{On the partition sum of the NS
  five-brane}'',
\href{http://arXiv.org/abs/hep-th/0205281}{{\tt hep-th/0205281}}.

\bibitem{Tsimpis:2007sx}
D.~Tsimpis, ``{Fivebrane instantons and Calabi-Yau fourfolds with flux}'', {\em
  JHEP} {\bf 03} (2007) 099,
\href{http://arXiv.org/abs/hep-th/0701287}{{\tt hep-th/0701287}}.

\bibitem{Donagi:2010pd}
R.~Donagi and M.~Wijnholt, ``{MSW Instantons}'',
\href{http://arXiv.org/abs/1005.5391}{{\tt 1005.5391}}.

\bibitem{Alexandrov:2010ca}
S.~Alexandrov, D.~Persson, and B.~Pioline, ``{Fivebrane instantons, topological
  wave functions and hypermultiplet moduli spaces}'', {\em JHEP} {\bf 03}
  (2011) 111,
\href{http://arXiv.org/abs/1010.5792}{{\tt 1010.5792}}.

\bibitem{Avramis:2006nb}
S.~D. Avramis, ``{Anomaly-free supergravities in six dimensions}'',
\href{http://arXiv.org/abs/hep-th/0611133}{{\tt hep-th/0611133}}.

\bibitem{Taylor:2011wt}
W.~Taylor, ``{TASI Lectures on Supergravity and String Vacua in Various
  Dimensions}'',
\href{http://arXiv.org/abs/1104.2051}{{\tt 1104.2051}}.

\bibitem{hopkins-2005-70}
M.~J. Hopkins and I.~M. Singer, ``Quadratic functions in geometry, topology,
  and {M}-theory'', {\em J. Diff. Geom.} {\bf 70} (2005) 329.

\bibitem{MR853982}
J.-M. Bismut and D.~S. Freed, ``The analysis of elliptic families. {I}.
  {M}etrics and connections on determinant bundles'', {\em Comm. Math. Phys.}
  {\bf 106} (1986), no.~1, 159--176.

\bibitem{MR861886}
J.-M. Bismut and D.~S. Freed, ``The analysis of elliptic families. {II}.
  {D}irac operators, eta invariants, and the holonomy theorem'', {\em Comm.
  Math. Phys.} {\bf 107} (1986), no.~1, 103--163.

\bibitem{Lee1988}
R.~Lee, E.~Y. Miller, and S.~H. Weintraub, ``Rochlin invariants, theta
  functions and the holonomy of some determinant line bundles'', {\em J. reine
  und angew. Math.} {\bf 392} (1988) 187--218.

\bibitem{AlvarezGaume:1983ig}
L.~Alvarez-Gaume and E.~Witten, ``{Gravitational Anomalies}'', {\em Nucl.
  Phys.} {\bf B234} (1984)
269.

\bibitem{Alvarez:1984yi}
O.~Alvarez, I.~M. Singer, and B.~Zumino, ``Gravitational anomalies and the
  family's index theorem'', {\em Commun. Math. Phys.} {\bf 96} (1984)
409.

\bibitem{AlvarezGaume:1984dr}
L.~Alvarez-Gaume and P.~H. Ginsparg, ``{The Structure of Gauge and
  Gravitational Anomalies}'', {\em Ann. Phys.} {\bf 161} (1985)
423.

\bibitem{AlvarezGaume:1983cs}
L.~Alvarez-Gaume and P.~H. Ginsparg, ``{The Topological Meaning of Nonabelian
  Anomalies}'', {\em Nucl. Phys.} {\bf B243} (1984)
449.

\bibitem{PhysRevLett.63.728}
F.~Bastianelli and P.~van Nieuwenhuizen, ``Gravitational anomalies from the
  action for self-dual antisymmetric tensor fields in 4k+2 dimensions'', {\em
  Phys. Rev. Lett.} {\bf 63} (Aug, 1989) 728--730.

\bibitem{Henneaux:1988gg}
M.~Henneaux and C.~Teitelboim, ``Dynamics of chiral (selfdual) p-forms'', {\em
  Phys. Lett.} {\bf B206} (1988)
650.

\bibitem{Belov:2006jd}
D.~Belov and G.~W. Moore, ``{Holographic action for the self-dual field}'',
\href{http://arXiv.org/abs/hep-th/0605038}{{\tt hep-th/0605038}}.

\bibitem{Monnier:2010ww}
S.~Monnier, ``{Geometric quantization and the metric dependence of the
  self-dual field theory}'',
\href{http://arXiv.org/abs/1011.5890}{{\tt 1011.5890}}.

\bibitem{MR0397797}
M.~F. Atiyah, V.~K. Patodi, and I.~M. Singer, ``Spectral asymmetry and
  {R}iemannian geometry. {I}'', {\em Math. Proc. Cambridge Philos. Soc.} {\bf
  77} (1975) 43--69.

\bibitem{Freed:1986zx}
D.~S. Freed, ``Determinants, torsion, and strings'', {\em Commun. Math. Phys.}
  {\bf 107} (1986)
483--513.

\bibitem{MR936082}
D.~S. Freed, ``{${\bf Z}/k$}-manifolds and families of {D}irac operators'',
  {\em Invent. Math.} {\bf 92} (1988), no.~2, 243--254.

\bibitem{Sati:2011pg}
H.~Sati, ``{On global anomalies in type IIB string theory}'',
  \href{http://arXiv.org/abs/1109.4385}{{\tt 1109.4385}}.

\bibitem{MR1785408}
E.~Rubei, ``Lazzeri's {J}acobian of oriented compact {R}iemannian manifolds'',
  {\em Ark. Mat.} {\bf 38} (2000), no.~2, 381--397.

\bibitem{MR1713785}
C.~Birkenhake and H.~Lange, {\em Complex tori}, vol.~177 of {\em Progress in
  Mathematics}.
\newblock Birkh\"auser Boston Inc., Boston, MA, 1999.

\bibitem{MR2062673}
C.~Birkenhake and H.~Lange, {\em Complex abelian varieties}, vol.~302 of {\em
  Grundlehren der Mathematischen Wissenschaften}.
\newblock Springer-Verlag, Berlin, second~ed., 2004.

\bibitem{2009arXiv0908.0555P}
A.~{Putman}, ``{The Picard group of the moduli space of curves with level
  structures}'', \href{http://arXiv.org/abs/0908.0555}{{\tt 0908.0555}}.

\bibitem{Sato01012010}
M.~Sato, ``The abelianization of the level 2 mapping class group'',
  \href{http://arXiv.org/abs/0804.4789}{{\tt 0804.4789}}.

\bibitem{Igusa1964}
J.-I. Igusa, ``On the graded ring of theta-constants'', {\em American Journal
  of Mathematics} {\bf 86} (1964) 219--246.

\bibitem{springerlink:10.1007/BF01231183}
D.~Johnson and J.~J. Millson, ``Modular {L}agrangians and the theta
  multiplier'', {\em Inventiones Mathematicae} {\bf 100} (1990) 143--165.

\bibitem{Brumfiel1973}
G.~W. Brumfiel and J.~W. Morgan, ``Quadratic functions, the index modulo 8 and
  a {Z}/4-{H}irzebruch formula'', {\em Topology} {\bf 12} (1973) 105--122.

\bibitem{Taylor1984259}
L.~R. Taylor, ``Relative rochlin invariants'', {\em Topology and its
  Applications} {\bf 18} (1984), no.~2-3, 259 -- 280.

\bibitem{MR0108467}
F.~van~der Blij, ``An invariant of quadratic forms mod {$8$}'', {\em Nederl.
  Akad. Wetensch. Proc. Ser. A 62 = Indag. Math.} {\bf 21} (1959) 291--293.

\bibitem{MR0322888}
E.~H. Brown, Jr., ``The {K}ervaire invariant of a manifold'', in {\em Algebraic
  topology ({P}roc. {S}ympos. {P}ure {M}ath., {V}ol. {XXII}, {U}niv.
  {W}isconsin, {M}adison, {W}is., 1970)}, pp.~65--71.
\newblock Amer. Math. Soc., Providence, R. I., 1971.

\bibitem{Snaith}
V.~Snaith, ``The {A}rf-{K}ervaire invariant of framed manifolds'',
  \href{http://arXiv.org/abs/1001.4751}{{\tt 1001.4751}}.

\bibitem{Freed:2006yc}
D.~S. Freed, G.~W. Moore, and G.~Segal, ``{Heisenberg groups and noncommutative
  fluxes}'', {\em Annals Phys.} {\bf 322} (2007) 236--285,
\href{http://arXiv.org/abs/hep-th/0605200}{{\tt hep-th/0605200}}.

\bibitem{Henningson:1999dm}
M.~Henningson, B.~E.~W. Nilsson, and P.~Salomonson, ``{Holomorphic
  factorization of correlation functions in (4k+2)-dimensional (2k)-form gauge
  theory}'', {\em JHEP} {\bf 09} (1999) 008,
\href{http://arXiv.org/abs/hep-th/9908107}{{\tt hep-th/9908107}}.

\bibitem{Witten:1999vg}
E.~Witten, ``Duality relations among topological effects in string theory'',
  {\em JHEP} {\bf 05} (2000) 031,
\href{http://arXiv.org/abs/hep-th/9912086}{{\tt hep-th/9912086}}.

\bibitem{Henningson:2010rc}
M.~Henningson, ``The partition bundle of type ${A}_{N-1}$ (2, 0) theory'', {\em
  JHEP} {\bf 04} (2011) 090,
\href{http://arXiv.org/abs/1012.4299}{{\tt 1012.4299}}.

\bibitem{Hatcher_2002}
A.~Hatcher, {\em Algebraic Topology}, vol.~227.
\newblock Cambridge University Press, 2002.

\bibitem{2010arXiv1003.5084C}
K.~{Cieliebak} and E.~{Volkov}, ``{First steps in stable Hamiltonian
  topology}'', \href{http://arXiv.org/abs/1003.5084}{{\tt 1003.5084}}.

\end{thebibliography}

\providecommand{\href}[2]{#2}\begingroup\raggedright\endgroup

}

\end{document}